\documentclass[10pt,twocolumn,floatfix,prb,superscriptaddress]{revtex4-2}
\usepackage{amsmath,amssymb,amsthm,mathrsfs,amsfonts,dsfont,amstext} 
\usepackage{textcomp,pbox}
\usepackage[export]{adjustbox}
\usepackage{bm}
\usepackage{dcolumn,booktabs,url}
\usepackage[scaled]{helvet}
\usepackage{sansmath,gensymb}
\usepackage{tikz,graphicx,transparent,color}
\usepackage{multirow}
\usepackage[separate-uncertainty = true]{siunitx}
\usepackage{comment}
\usepackage{physics}
\usepackage{pdfpages}
\usepackage{array}
\usepackage{makecell}
\usepackage{dsfont}
\usepackage{tabularx}
\newcolumntype{L}{>{\raggedright\arraybackslash}X}
\newcolumntype{Y}{>{\centering\arraybackslash}X}

\makeatletter
\AtBeginDocument{\let\LS@rot\@undefined}
\makeatother

\newcolumntype{C}[1]{>{\centering\let\newline\\\arraybackslash\hspace{0pt}}m{#1}}

\newcommand{\red}[1]{\textcolor{red}{#1}}

\usepackage{natbib} 

\usepackage[colorlinks=true]{hyperref}

\usepackage{mathtools} 

\hypersetup{
     colorlinks   = true,
     citecolor    = red,
     linkcolor    = blue,
     urlcolor     = red     
}

\graphicspath{{./figs/}}

\newcommand{\figref}[1]{\figurename{~\ref{#1}}}
\newcommand{\tabref}[1]{\tablename{~\ref{#1}}}

\newcommand{\transition}{$^2$F$_{7/2}(0) \longleftrightarrow ^2$F$_{5/2}$(0)  }

\newcommand{\dfc}{$^2$F$_{5/2}$}
\newcommand{\dfs}{$^2$F$_{7/2}$}

\newcommand{\yb}{{Yb$^{3+}$}}
\newcommand{\ybiso}{{$^{171}$Yb$^{3+}$}}
\newcommand{\ybCWO}{{$^{171}$Yb$^{3+}$:CaWO$_4$}}
\newcommand{\ybYSO}{{$^{171}$Yb$^{3+}$:Y$_2$SiO$_5$}}
\newcommand{\euYSO}{{Eu$^{3+}$:Y$_2$SiO$_5$}}
\newcommand{\CWO}{{CaWO$_4$}}

\usepackage{lipsum}
\usepackage{graphicx}

\usepackage{import}
\usepackage{xifthen}
\usepackage{pdfpages}
\usepackage{transparent}
\usepackage{svg}
\usepackage{adjustbox}
\usepackage{calc}
\graphicspath{{./figs/}}
\usepackage{physics}

\makeatletter
\def\maketitle{
\@author@finish
\title@column\titleblock@produce
\suppressfloats[t]}
\makeatother

\usepackage[resetlabels]{multibib}

\newcites{S}{References}

\begin{document}
\newcommand{\TitleName}{Sub-second spin and lifetime-limited optical coherences in \ybCWO{} }

\title{\TitleName}

\newcommand{\AffChimieParis}{Chimie ParisTech, PSL University, CNRS, Institut de Recherche de Chimie Paris, 75005 Paris, France }
\newcommand{\AffCaltech}{Thomas J. Watson, Sr., Laboratory of Applied Physics, California Institute of Technology, Pasadena, CA, 91125, USA}
\newcommand{\AffIMNP}{CNRS, Aix Marseille Univ, Université de Toulon, Institut Matériaux Microélectronique et Nanosciences de Provence, IM2NP, 13013 Marseille, France}

\author{Alexey Tiranov$^{\dag}$}
\thanks{Present address: Sparrow Quantum ApS, 2000 Frederiskberg, Denmark}
\affiliation{\AffChimieParis{}}
\author{Emanuel Green}
\affiliation{\AffCaltech{}}
\author{Sophie Hermans}
\affiliation{\AffCaltech{}}
\thanks{Present address: Delft}
\author{Erin Liu}
\affiliation{\AffCaltech{}}
\author{Federico Chiossi}
\affiliation{\AffChimieParis{}}
\thanks{Present address: Padova}
\author{Diana Serrano}
\affiliation{\AffChimieParis{}}
\author{Pascal Loiseau}
\affiliation{\AffChimieParis{}}
\author{Achuthan Manoj Kumar}
\affiliation{\AffIMNP{}}
\author{Sylvain Bertaina}
\affiliation{\AffIMNP{}}
\author{Andrei Faraon}
\affiliation{\AffCaltech{}}
\author{Philippe Goldner}
\thanks{alti@sparrowquantum.com, philippe.goldner@chimieparistech.psl.eu}
\affiliation{\AffChimieParis{}}

\date{\today}

\begin{abstract}

Optically addressable solid-state spins have been extensively studied for quantum technologies, offering unique advantages for quantum computing, communication, and sensing. Advancing these applications is generally limited by finding materials that simultaneously provide lifetime-limited optical and long spin coherences. Here, we introduce \ybiso{} ions doped into a \CWO{} crystal. We perform high-resolution spectroscopy of the excited state, and demonstrate all-optical coherent control of the electron-nuclear spin ensemble. We find narrow inhomogeneous broadening of the optical transitions of 185 MHz and radiative-lifetime-limited coherence time up to 0.75~ms. Next to this, we measure a spin-transition ensemble line width of 5 kHz and electron-nuclear spin coherence time reaching 0.15~seconds at zero magnetic field between 50~mK and 1~K temperatures. These results demonstrate the potential of \ybCWO{} as a low-noise platform for building quantum technologies with ensemble-based memories, microwave-to-optical transducers, and optically addressable single-ion spin qubits.

\end{abstract}

\maketitle 

Solid-state optically addressable spins \cite{Aharonovich2016,Atature2018,Awschalom2018} have been used to demonstrate a variety of applications including quantum state storage \cite{Bussires2013,Jobez2015,Gundogan2015}, spin-spin entanglement \cite{Bernien2013}, control of nuclear spin clusters \cite{Bradley2019}, and quantum sensing \cite{Taylor2008}. Among the most popular platforms are optically active defects in diamond \cite{Chu2014}, semiconductor quantum dots \cite{Lodahl2015}, vacancies in silicon \cite{Redjem2020}, silicon carbide \cite{Morioka2020} and hexagonal boron nitride \cite{Caldwell2019}. Despite significant progress, the search for defects in a solid-state material simultaneously exhibiting lifetime-limited optical and long spin coherences remains ongoing. 

Rare-earth (RE) ion based defects are particularly attractive due to their unique electronic configuration, leading to narrow optical transitions and long coherence times in many different crystalline host materials \cite{Tittel2010,Rancic2017,Kindem2018,Ortu2018,Welinski2020,LeDantec2021,Nicolas2023}. Ensembles of such defects have been used for quantum memories \cite{Bussires2013}. Ensembles of rare-earth ions have  been extensively explored as quantum memories for photonic \cite{Bussires2013} and microwave qubits \cite{Grezes2016} and optical-to-microwave transduction \cite{fernandez-gonzalvo_coherent_2015,Bartholomew2020}. Single emitters coupled to integrated nanophotonic cavities \cite{Dibos2018,Zhong2018,Ulanowski2022,Yu2023,Huang2023,Yang2023} are actively explored as they enable  single-shot spin readout \cite{Raha2020,Kindem2020,Gritsch2025}, addressing of  nearby nuclear spins \cite{Kornher2020,Ruskuc2022,Uysal2023}, spin-photon entanglement \cite{Uysal2025}, and  entanglement between remote ions \cite{Ruskuc2025}.

\CWO{} has recently drawn a lot of attention as an appealing  host material because of its low magnetic impurity background compared to conventional yttrium-containing hosts. The remaining magnetic background is dominated by the $^{183}$W isotope (low natural abundance of 14.3\%) and paramagnetic impurities such as iron and other RE ions \cite{Billaud2025}. The non-polar defect site symmetry yields a first-order insensitivity to electric fields. These material properties combined have enabled the demonstration of spin coherence times up to 23 ms for small ensembles ($<$1 ppb) in Er$^{3+}$:CaWO$_4$ \cite{LeDantec2021} as well as two-photon interference and
spin-photon entanglement using a single Er$^{3+}$ ion in CaWO$_4$ coupled to a heterogeneous photonic crystal cavity \cite{Uysal2025}.

In this work, we investigate a previously unexplored host-ion combination: \ybiso{} doped into \CWO{}. Our results establish \ybCWO{} as a unique material platform possessing radiatively limited optical coherences and sub-second electron-nuclear spin coherence, even at appreciable doping concentrations.
Among the trivalent RE species, \ybiso{} is the only paramagnetic isotope to possess a nuclear spin of $I = 1/2$. The hybrid electron-nuclear spin states can be used as highly-coherent states, while a low $I$ ensures simple spin manipulation. 
Using high-resolution optical spectroscopy we extract the so-far unknown excited state Hamiltonian parameters of \ybiso{} in \CWO{}. Having complete knowledge of the energy levels and the optical transitions allows us to realize efficient all-optical spin initialization and control. In ensembles, we demonstrate optical coherence times  reaching $0.75$~ms, nearly limited by radiative decay, and realize all-optical spin control with spin coherence times up to 0.15(1)~s. Coherence properties are achieved at zero magnetic field and with temperatures up to a few K. 
Compared to other solid-state emitters possessing electron spins including defects in diamond, silicon and  and other RE ion-doped systems, our results represent a substantial advance in both optical and spin coherence times,  even at elevated dopant concentrations.

\begin{figure}[hbtp!]
	\includegraphics[width=1\linewidth, trim=0.cm 0.05cm 0.cm 0.0cm,clip]{./figures/figure1.pdf}
	\caption{ 
	(color online) \textbf{Energy level diagram and magnetic field dependence of \ybCWO{}.}
    (a) \CWO{} crystal structure, with a  \ybiso{} dopant replacing a Ca$^{2+}$ ion at a site of $S_4$ point symmetry.
    (b) High-resolution absorption spectra of \ybCWO{} at 4~K for light electric field polarized along the $c$ axis ($E \parallel c$) or perpendicular ($E \perp c$) to the $c$ axis. The labeling of the absorption peaks corresponds to the energy level diagram (c). Absorption peak C corresponds to  Yb$^{3+}$ isotopes with zero nuclear spin. Absorption peak D only appears for $E \perp c$.
    (c) Low-field energy level diagram for the $^2$F$_{7/2}(0) \rightarrow ^2$F$_{5/2}(0)$ transition of \ybCWO{} at 973.1 nm. Energy splittings in the ground and excited state are determined using the ground state and extracted excited-state hyperfine tensors. The transitions corresponding to the observed absorption spectrum in (b)  are shown with solid lines.
    (d) Absorption spectra of \ybCWO{} for varying magnetic field strengths applied perpendicular to the crystalline $c$-axis. The polarization of the incident light is $E \perp c$. Dashed lines show the results of the fit to the model given by Eq. (\ref{eq:Heff}). Purple lines denote energy levels of the \ybiso{} ions and blue lines denote energy levels of the $I=0$ isotopes. The color scale is linear.
}
	\label{fig:1}
\end{figure}

\vspace{5pt}
\textbf{High-resolution spectroscopy of the ground and excited level structure}.
A \CWO{} crystal doped with \ybiso{} ions was grown using the Czochralski method with a nominal concentration of 20~ppm~\cite{SM}. In this work, we study the tetragonal $S_4$ site in which the \yb{} replaces a Ca$^{2+}$ (\figref{fig:1}(a)). This dopant's optical transition connecting \dfs{}(0) $\rightarrow$ \dfc{}(0) is centered around 973.16 nm (vac.) 
\cite{SM}. The concentration of ions in the $S_4$ site was measured to be $\approx 5$ ppm using electron paramagnetic resonance, with an isotopic purity of 96\%. The high-resolution optical absorption spectrum at 4 K 
showed a series of narrow peaks spanning several GHz (\figref{fig:1}(b)). 
The narrow optical inhomogeneous broadening of 185~MHz indicates the high quality of the material and low distortion of the crystalline site induced by dopant ions. This allows us to optically address each ground state level separately (\figref{fig:1}(b)).

\ybiso{} carries an electron spin $S = 1/2$  and nuclear spin $I=1/2$ , and both its ground and excited states spin levels can be described using the effective Hamiltonian  
\begin{equation}
\label{eq:Heff}
\mathcal{H}^{g/e} = \mathbf{I} \cdot \mathbf{A}^{g/e} \cdot \mathbf{S} + \mu_\text{B} \mathbf{B} \cdot \mathbf{g}^{g/e} \cdot \mathbf{S} - \mu_\text{n}  \mathbf{B}\cdot \mathbf{g}_\text{n} \cdot \mathbf{I},
\end{equation}
where $\mathbf{g}^{g/e}$, $\mathbf{g}_\text{n}$ and $\mathbf{A}^{g/e}$ are the coupling tensors of the electronic  Zeeman, nuclear Zeeman, and hyperfine interactions, respectively, and $\mu_B$ and $\mu_n$ are the electronic and nuclear spin magnetons. The superscript indicates the ground or excited state manifold.
The $S_4$ point symmetry of the tetragonal occupation site significantly simplifies the spin interaction, giving a symmetry axis parallel to the $c$-axis of the crystal. As a result, the uniaxial interaction tensors consist of one parallel and two identical perpendicular components $A_{\parallel}^{g/e}$, $A_{\perp}^{g/e}$ and $g_{\parallel}^{g/e}$, $g_{\perp}^{g/e}$, for the $\mathbf{A}$ and $\mathbf{g}$ tensors, respectively. The nuclear Zeeman interaction $\mathbf{g}_\text{n}$ tensor is considered to be isotropic with $g_n = 0.987$ for \ybiso{} nuclear spin \cite{Stone2005}.

We identified the energy level diagram for the \dfs{}(0) and \dfc{}(0) levels using the high-resolution optical spectroscopy with  different polarisations of the incident light (\figref{fig:1}(c)). We extract the hyperfine tensors in the ground \cite{Rakhmatullin2009} and excited states: $A^g_{\parallel}/h = 0.787(1)$ GHz, $A^g_{\perp}/h = 3.08384(1)$ GHz and $A^e_{\parallel}/h = -2.878(2)$ GHz, $A^e_{\perp}/h = 2.734(2) $ GHz. 
To complete the optical spectroscopy, we determine the excited state $g^e$-tensor by  fitting the magnetic field dependence of the energy levels to the model given by Eq.~\eqref{eq:Heff}. We sweep the  magnetic field by varying the current through a set of Helmholtz coils mounted parallel (perpendicular) to the crystalline c-axis (\figref{fig:1}(d)). The extracted values  are $g^e_{\parallel} = -1.452(1)$ and $g^e_{\perp} = 1.362(1)$, qualitatively matching previous results \cite{Jones1967}. At non-zero magnetic fields, symmetry-forbidden optical transitions become apparent as the symmetry of the site is reduced and and additional spin mixing occurs due to the combined hyperfine and Zeeman interactions, enabling transitions that are forbidden at zero field.
Surprisingly, to fully explain the optical absorption spectrum, we need to assume higher $D_{2d}$ point symmetry, leading to $S_4$ site symmetry under a perturbation \cite{SM,Hull2005}.

\begin{figure}[hbtp!]
	\includegraphics[width=0.99\linewidth, trim=0.cm 0.05cm 0.cm 0.0cm,clip]{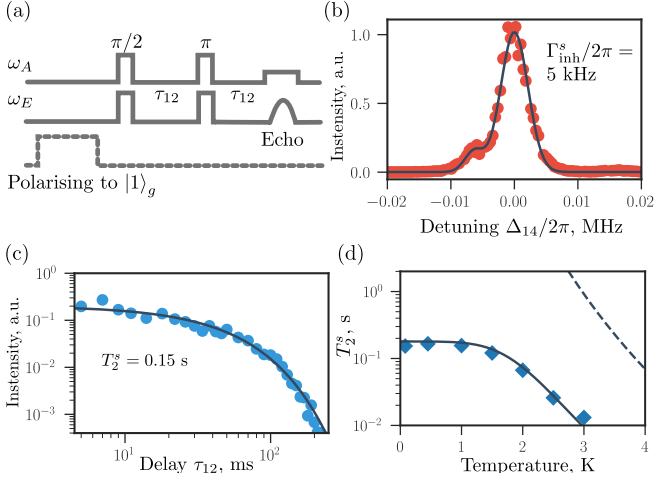}
	\caption{ 
	(color online) \textbf{All-optical coherent spin control.}
    (a) Experimental pulse sequence  used for all-optical detection of spin coherence. First, we polarise the spin ensemble of \ybiso{} ions into the $\ket{1}_g$ state by optical pumping around the $\omega_A$ and $\omega_B$  frequencies. We then use optical pulses with frequency $\omega_A$ and $\omega_E$ to induce a Raman transition between the $\ket{1}_g$ and $\ket{4}_g$ spin states. The spin echo sequence with $\tau_{12}$ delay between the first two spin rotations ends with a pulse at the $\omega_A$ frequency used for detecting the spin echo signal. 
    (b) Measurement of the $\ket{1}_g$ - $\ket{4}_g$ transition illustrating the inhomogeneous broadening at zero magnetic field with $\Delta_{14}/2\pi$ detuning around $\omega_A-\omega_E  = 3083.85$~MHz. A Gaussian fit gives a FWHM of $\Gamma^s_{\text{inh}}/2\pi$= 5(1) kHz.
    The  shoulder on the main peak is presumed to arise from fast spin flip-flop dynamics depending  on the detuning, which modifies the measured lineshape; further measurements at different doping concentrations are needed to confirm this effect.
    (c) Spin echo intensity as a function of the delay $\tau_{12}$ between the two spin rotations, at zero magnetic field. The solid line is the fit giving $T^s_2 = 0.15(1)$~s.
    (d) Spin coherence time as a function of the temperature, explained by the model based on flip-flops. The dashed line represents the measured $2T_1$ limit given by the spin-lattice interaction \cite{SM}. The spin echo decays were fitted with a pure exponential function; no stretch factor was needed, suggesting that the dominant dephasing mechanism remains spin flip-flop dynamics.
}
	\label{fig:2}
\end{figure}

At zero magnetic field, the eigenstates of the ground state manifold can be expressed as  
\begin{equation}
\begin{split}
  \ket{1}_g &= (\ket{\uparrow \Downarrow }_g - \ket{\downarrow \Uparrow }_g)/\sqrt{2}, \\ 
  \ket{2,3}_g &= \ket{\uparrow \Uparrow }_g, \ket{\downarrow \Downarrow }_g, \\
  \ket{4}_g &= (\ket{\uparrow \Downarrow  }_g + \ket{\downarrow \Uparrow }_g )/\sqrt{2},
\end{split}
\end{equation}
and the excited state manifold as 
\begin{equation}
\begin{split}
  \ket{1,2}_e &= \ket{\uparrow \Uparrow }_e, \ket{\downarrow \Downarrow }_e, \\ 
  \ket{3}_e &= (\ket{\uparrow \Downarrow }_e + \ket{\downarrow \Uparrow }_e)/\sqrt{2}, \\
  \ket{4}_e &= (\ket{\uparrow \Downarrow  }_e - \ket{\downarrow \Uparrow }_e)/\sqrt{2}, 
\end{split}
\end{equation}
where $\ket{\uparrow} = \ket{m_S=+1/2}$, $\ket{\downarrow} = \ket{m_S=-1/2}$ represent the electron spin and $\ket{\Uparrow} = \ket{m_I=+1/2}$, $\ket{\Downarrow} = \ket{m_I=-1/2}$ the nuclear spin projections. 
Two out of four electron-nuclear hyperfine states are completely entangled in their electronic and nuclear components (\figref{fig:1}(a)). The average magnetic moment for these states is zero ($\bra{i}\mathbf{S}\ket{i} = \bra{i}\mathbf{I}\ket{i} = 0$), rendering the energy levels insensitive to external magnetic fields in  first order. As a result, the $\ket{1}_g - \ket{4}_g$ spin transition connecting two hybridized states is considered a clock transition. Similarly, optical transitions such as $\ket{4}_g - \ket{4}_e$ are also first-order protected from magnetic noise (\figref{fig:1}(a)).

The combination of large hyperfine splittings and narrow optical lines allows for optical initialization of the whole spin ensemble into a desired ground state level. By addressing the $\ket{4}_g$ and $\ket{2,3}_g$ states via the A and B transitions (\figref{fig:1}(b)), we prepare \ybiso{} ions into the  $\ket{1}_g$ state with high efficiency. Efficient optical pumping is possible thanks to strongly suppressed spin-phonon relaxation processes at temperatures below 2~K. We measure a spin ensemble recovery time of more than 3 hours at 50 mK \cite{SM}. Together with a relatively short excited state lifetime of $T_{1}^{o} = 0.385$ ms, we achieve more than 99\% initialization efficiency for the whole \ybiso{} spin ensemble under the experimental conditions specified in \cite{SM} (Section IV.D), where optical pumping was performed repeatedly until a steady-state population distribution was reached.

\vspace{5pt}
\textbf{Spin coherence}.
Next, we implement all-optical coherent spin control to measure the spin coherence properties of the \ybiso{} ensemble. We use an optical $\Lambda$ system, consisting of transitions A and E, to address the $\ket{1}_g - \ket{4}_g$ spin transition at 3.083 GHz (\figref{fig:1}(a,b)).  To drive the Raman transition, we use an electro-optic modulator to generate two spectral tones separated by the  $\ket{1}_g - \ket{4}_g$ energy splitting. 
Using this control, we perform a spin echo experiment consisting of a sequence of $\pi/2$ and $\pi$ pulses separated by a $\tau_{12}$ delay (\figref{fig:2}(a)). With a final probe pulse addressing the  $\ket{4}_g$ state using the A transition, we observed the spin echo by detection of the coherent Raman scattering \cite{GuillotNol2009}. First, we varied the frequency separation of the two laser tones and detected the spin echo signal at a fixed delay to extract a spin inhomogeneous linewidth of 5(1) kHz (\figref{fig:2}(b)). The observed spin inhomogeneous broadening is one of the narrowest reported for paramagnetic solid-state ensembles, less than the 48~kHz in $^{171}$Yb$^{3+}$:YVO$_4$ \cite{Kindem2020} and 0.5~MHz in $^{171}$Yb$^{3+}$:Y$_2$SiO$_5$ \cite{Tiranov2018}. Then, by varying the delay between two pulses, we measured a spin coherence time up to $T_{2}^{s} = 0.15(1)$ s (\figref{fig:2}(c,d)). The measured spin coherence time is the longest achieved at zero magnetic fields among RE ions, even those possessing nuclear spin (15~ms in $^{151}$Eu$^{3+}$:Y$_2$SiO$_5$ \cite{Alexander2007}).

\begin{figure*}[hbtp!]
	\includegraphics[width=0.99\linewidth, trim=0.cm 0.02cm 0.cm 0.0cm,clip]{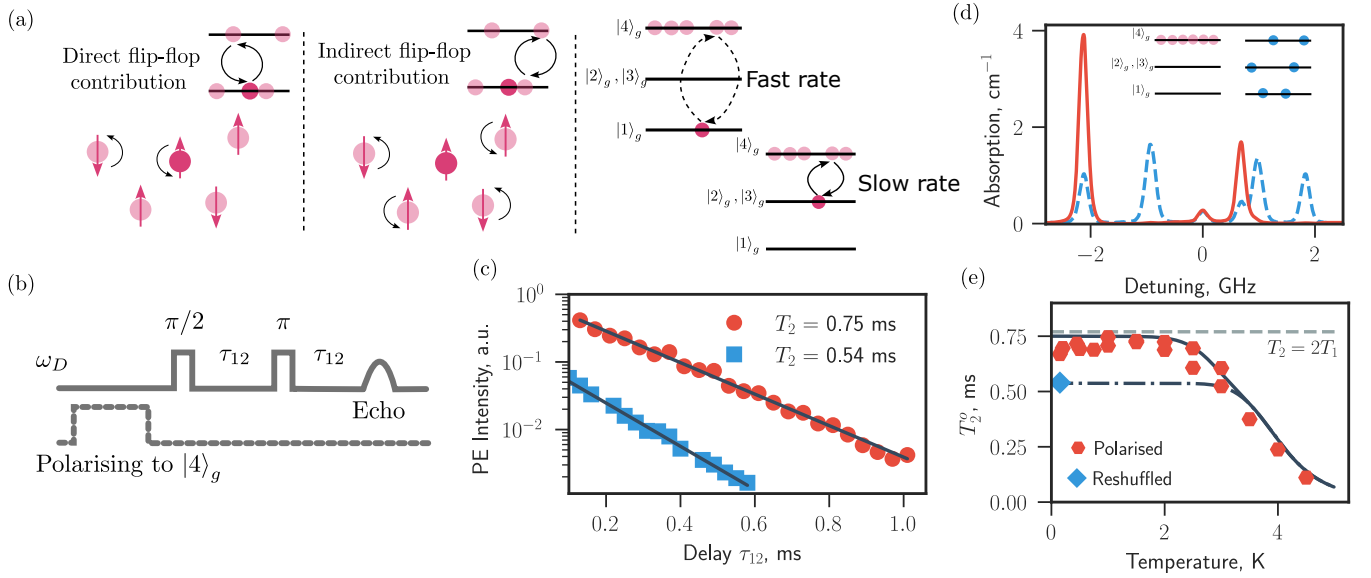}
	\caption{ 
	(color online) \textbf{Optical coherence of \ybCWO{}.}
    (a) Left: flip-flops between the addressed ions and the neighboring spins directly contributes to a reduction in spin coherence. Middle: indirect contribution to decoherence due to flips of neighboring spin pairs (light red) producing magnetic noise for the probed ions (dark red).
    Right: spin dynamics rates for different transitions are highly transition dependent. Flip-flop rates with states involving $\ket{2,3}_g$ are two orders of magnitude larger than for $\ket{1}_g$ -  $\ket{4}_g$  transition.
    (b) Experimental pulse sequence used for photon echo (PE) measurements. It consists of spin polarisation by optical pumping using $\omega_E$ and $\omega_F$  frequencies (see \figref{fig:1}(a)). 
    Optical pulses resonant with the D transition are then used for the photon echo (PE) sequence with a $\tau_{12}$ delay between the two pulses. 
    (c) Measurement of the PE intensity at the lowest temperature (50 mK) as a function of the delay $\tau_{12}$ after polarising the spin ensemble of \ybiso{} ions into $\ket{4}_g$ state (red) or reshuffling population between all the spin states (blue). Solid lines are exponential fits, giving optical coherence times $T^o_2 = 0.75$~ms and $T^o_2 = 0.54$~ms, respectively. 
    (d) Absorption profile after polarising the spin ensemble into $\ket{4}_g$ state (solid line) or reshuffling population between all the spin states (dashed line).
    (e) Optical coherence time as a function of temperature while reshuffling the spin populations (blue diamond). The optical coherence time when polarising spins into the $\ket{4}_g$ state is enhanced (red circles) and approaches the $2T_1$ limit (dashed line). The solid and dash-dot lines show predictions from the model \cite{SM}.
}
	\label{fig:pe}
\end{figure*}

This long spin coherence time is governed by several features. First, the nuclear spin density of \CWO{} is one of the lowest for  crystalline materials and is dominated by the $^{183}$W isotope \cite{Kanai2022}. Additionally,  $^{183}$W has a rather low $g_n = 0.2356$, resulting in weak interactions with \ybiso{} spins. Second, the $\ket{1}_g - \ket{4}_g$ transition is protected by the clock condition, i.e. it is only second-order sensitive to external magnetic noise. Third, the spin-spin dynamics between  \ybiso{} ions are suppressed due to the polarisation into the $\ket{1}_g$ state. We measured the temperature dependence of the coherence time up to 3~K, while repeating the spin polarisation in $\ket{1}_g$ via optical pumping (\figref{fig:2}(d)). The measured spin coherence time is relatively constant up to 1 K and decreases to 10 ms at 3 K. We attribute this drop to \ybiso{} spins experiencing flip-flop processes that exchange their spin states due to long-range magnetic dipole-dipole interaction \cite{Welinski2020}.
More specifically, at 1K, the optical pumping efficiency starts to drop due to the stronger spin-phonon relaxation and prevents the complete emptying of the $\ket{2,3}_g$ states. At 3~K, the spin-lattice relaxation rate increase to $\geq$2 s$^{-1}$~\cite{SM}, resulting in $\approx 10$\% of population staying in $\ket{2,3}_g$ states after the initial polarisation.
This intensified flip-flop process between \ybiso{} ions  affects the spin coherence in several ways (\figref{fig:pe}(a)) \cite{Tyryshkin2011}. The spin coherence is directly affected due to an increase in spin flip-flops involving spins in $\ket{2,3}_g$ state, which is efficiently quenched by the optical pumping at lower temperatures. Indirectly, the magnetic field noise is increased due to increased spin flip-flops of \ybiso{} spins near the probed ensemble. However, this effect is assumed to be small due to the low magnetic field sensitivity of the states and the modest concentration of \ybiso{} spins. 

\vspace{5pt}
\textbf{Lifetime-limited optical coherence}.
In this section, we measure the optical coherence, and subsequently use the optical coherence measurements to provide more insight into the fast spin-spin dynamics.  The optical coherence time $T_2^{o}$ of the $\ket{4}_g - \ket{4}_e$ transition is measured through a standard photon echo technique using a single laser tone resonant with the D transition (\figref{fig:pe}(a)). Without optical pumping, an optical coherence time $T_2^{o}$ up to 0.54 ms was measured, surpassing the excited state lifetime $T_1^o = 0.385$ ms (\figref{fig:pe}(b,c)). Next, using optical pumping, we initialize all the population in the $\ket{4}_g$ state (\figref{fig:pe}(d)). In this manner, the optical coherence time increases to $T_2^{o} = 0.75(2)$ ms, which is nearing the $2T_1 = 0.77$ ms limit (\figref{fig:pe}(e)).

We attribute the decrease of the optical coherence time without optical pumping to spin-spin dynamics via magnetic dipole-dipole interactions involving the $\ket{2,3}_g$ state, described above.  Due to the high anisotropy of the $\mathbf{g}$ tensor, the spin flip-flop processes are strongly state-dependent and the flip-flop rate is proportional to $g^4$ \cite{Welinski2020}.  Only the $\ket{1}_g - \ket{4}_g$ spin transition is solely coupled via $g_{\parallel}$ (since only $\bra{1}_g S_z \ket{4}_g$ is non-zero, see Section 5 in \cite{SM}), all other spin transitions involve  $g_{\perp}$ which is almost four times higher. Hence, the flip-flop rates involving the $\ket{2,3}_g$  are faster by a factor  $g_{\parallel}^4/g_{\perp}^4 \approx 200$ and greatly reduce the spin coherence time $T_2^s$ (\figref{fig:pe}(a)). We estimate the flip-flop rates involving the $\ket{2,3}_g$ states to reach $10^3$ s$^{-1}$, leading to millisecond spin lifetimes \cite{Welinski2020}. Removing the population from $\ket{2,3}_g$ states is thus crucial to suppress the direct contribution from spin flip-flop  dynamics (\figref{fig:pe}(a)). Above 2~K, the additional decrease of the optical coherence can be explained by phonon scattering in the excited state with a $T^9$ dependence (\figref{fig:pe}(e)).
This optical coherence time of 0.75~ms is one of the longest observed for paramagnetic solid-state emitters comparable to  \ybYSO{} (1~ms \cite{Chiossi2024}) and nuclear spin ions in \euYSO{}, (2.6~ms \cite{Equall1994}).

\vspace{5pt}
\textbf{Discussion and conclusion}.
We stress that even if the energy levels are insensitive to magnetic fields in the first order, the dipole moments of their transitions remain strong and correspond to electron spins. Such that, $\bra{1}_g \mu_\text{B} \mathbf{g} \cdot \mathbf{S} \ket{4}_g = \mu_\text{B} g_{\parallel}^g
$, the spin transition can be driven using an AC magnetic field along the $c$-axis. The optimal orientation of the AC drive for all other spin transitions is perpendicular to the $c$-axis, with the transition dipole moment proportional to $g_{\perp}^g$. We emphasize that even under the clock condition, microwave driving can still be efficient and fast, making this system compatible with interfacing with superconducting circuits \cite{LeDantec2021,Alexander2022}.

Due to these strong transition dipole moments, electron spin-like fast spin-spin dynamics are present between \ybiso{} ions. As we demonstrate in this study, such dynamics can be suppressed at low temperatures  by polarising the whole spin ensemble via optical pumping. However, applications involving large ensembles must have long-lived auxiliary spin states storing part of the population. These states must be free from fast spin-spin processes to prevent coupling the spin population to other states. One way to suppress the fast spin-spin dynamics is to work with lower doping concentrations. For example, lowering the concentration to 0.5~ppm will suppress the fast flip-flop rate down to $10^2$ s$^{-1}$ level, while still exhibiting significant optical absorption.  To reach the regime where decoherence is limited by the W nuclear spin bath, concentrations on the order of a few hundred ppb are expected to be required.
Another solution is to perform additional engineering of the material, and to controllably broaden the spin inhomogeneous linewidth without affecting the optical absorption. An example of this approach has been explored through co-doping of rare earth ions~\cite{Thiel2012,Welinski2017}.

Lowered doping concentrations would enable the observation of single \ybiso{} in \CWO{} and is expected to yield even longer spin and optical coherence times. Promisingly, a 0.7~ppb natural concentration of Er$^{3+}$ allowed for 23~ms spin coherence times at high magnetic fields \cite{LeDantec2021}. The use of zero-field clock transitions, reported in our work, will strongly enhance coherences, while integration with nanophotonic structures will improve the properties of a spin-photon interface \cite{Uysal2025,Ruskuc2025}.  Several photonic integration strategies are under development, including hybrid nanophotonic platforms where photonic structures are either transferred onto the bulk crystal \cite{Uysal2025} or directly fabricated within it \cite{Ruskuc2025}. Additionally, recent work on the growth of epitaxial \CWO{} thin films will enable future monolithic photonic and electronic integration \cite{Tang2024}.  Importantly, \CWO{} offers a unique advantage to realize a completely nuclear-spin-free environment, through proper isotopic purification of Ca and W.

In conclusion, we have experimentally investigated the novel material \ybCWO{} and reached lifetime-limited optical coherence and spin coherence times up to 0.15(1)~s, facilitated by the clock transitions at zero magnetic fields. 
This represents an order-of-magnitude increase compared to the previously studied \ybYSO{}, which exhibited spin coherence times up to 10 ms under similar conditions \cite{Nicolas2023}. The enhanced performance of \ybCWO{} can be attributed to the lower density of nuclear spins in the \CWO{} host crystal. Combined with a reduced concentration of \ybiso{}, this  suggests that even longer coherence times can be achieved than those measured in the present work.
This material provides a set of exclusive properties such as optically resolved hyperfine transitions, the ability for all-optical initialization of the spin ensemble, and long coherence times, making it a unique resource for quantum networking and information applications. These features make \ybCWO{} appealing for both ensemble-based applications, where collective ensemble coupling is beneficial, and integrated spin-photon interfaces based on single emitters. These applications include optical quantum memories \cite{Bussires2013,Businger2022}, coupling to superconducting circuits \cite{LeDantec2021},  integrated spin-photon interfaces \cite{Uysal2025,Ruskuc2025}, and optical-to-microwave quantum transducers \cite{Bartholomew2020}.
\newline\\
\textbf{Data availability.}
The data from the plots within this paper and other details of this study are available from the corresponding author upon request and are also openly available at \href{https://doi.org/10.6084/m9.figshare.30455660}{https://doi.org/10.6084/m9.figshare.30455660}.

\bibliographystyle{apsrev4-1} 

\let\oldaddcontentsline\addcontentsline
\renewcommand{\addcontentsline}[3]{}
\let\addcontentsline\oldaddcontentsline
\textbf{Acknowledgement.}
This project has received funding from the CNRS under the Priority Quantum Research Programme and Equipment (PEPR, QMemo) (P.G.), NASA ROSES program, Gordon and Betty Moore Foundation Experimental Physics Investigators program (A.F.). A.T. was supported by ANR under grant agreement no. ANR-22-CPJ2-0060-01. E.G. acknowledges support from the the National Gem Consortium as well as the National Science Foundation Graduate Research Fellowship under grant no. 2139433. S.H. acknowledges support from the AWS Quantum Postdoctoral Fellowship. E.L. acknowledges support from the DoD National Defense Science and Engineering Graduate Fellowship. S.B. thanks the support of the CNRS research infrastructure INFRANALYTICS (FR 2054). F.C. acknowledges financial support from Marie Curie Fellowship (101066781). A.M.K. is supported by ANR QuantEdu-France (22-CMAS-0001) and by France 2030 investment plan, as part of the Initiative d'Excellence d’Aix-Marseille Université – A*MIDEX (AMX-22-RE-AB-199).
Authors thank M. Afzelius, L. Nicolas for discussions, and A. Ferrier for technical help.
\newline\\
\textbf{Contributions.}
The crystal growth was done by A.T., P.L., and P.G. Initial optical spectroscopy was performed by A.T. and P.G. EPR measurements were performed by S.B. and A.M.K. Optical characterization under magnetic field was done by E.G., S.H., and E.L. with support from A.F. All the photon echo and optically detected spin echo measurements were carried out by A.T. with support from P.G. and results were interpreted by A.T. and P.G. with contributions from F.C. and D.S. The manuscript was mainly written by A.T. and P.G., with contributions from all the authors.
\newline
\textbf{Competing interests.}
The authors declare no competing interests.

\pagebreak
\clearpage 

\title{Supplementary Notes - \TitleName}

\maketitle
\onecolumngrid
\setcounter{equation}{0}
\setcounter{figure}{0}
\setcounter{table}{0}
\setcounter{page}{1}
\makeatletter
\renewcommand{\theequation}{S\arabic{equation}}
\renewcommand{\thefigure}{S\arabic{figure}}
\renewcommand{\thetable}{S\arabic{table}}
\renewcommand{\bibnumfmt}[1]{[S#1]}
\renewcommand{\citenumfont}[1]{S#1}
\renewcommand{\@seccntformat}[1]{%
  \csname the#1\endcsname.\quad
}

{
  \hypersetup{linkcolor=black}
  \tableofcontents
}

\setcounter{secnumdepth}{3} 
\newpage

\section{Sample preparation}
\label{SM:grow}

We used \CWO{} scheelite crystals nominally doped with 0.002 at.\% (20 ppm) \ybiso{} ions. The crystal sample was cut from a boule grown using the Czochralski method, starting from  CaCO$_3$ and WO$_3$ raw materials with 99.97 at.\% and 99.98 at.\% purity, respectively. 
Isotopically purified $^{171}$Yb$_2$O$_3$  with 95\% purity was used for the doping. 

Samples were cut along the $a$ ($b$) and $c$ crystallographic axes. Several samples with 1 cm and 1.5 cm sizes along the $c$ axis were prepared, and polishing was done to get optically transmitting surfaces along several directions. For optical experiments, two samples with different sizes (9.2, 9.7, 9.9) and (15, 4.1, 4.8) mm in the ($a$, $a$, $c$) frame were prepared. For the electron paramagnetic resonance (EPR) study, a (1.5, 2.2, 2.5) mm sample weighing 63 mg was used.\\

\section{Crystal structure}

\CWO{} has a tetragonal structure belonging to the $I4_1/a$ space group (\#88). \yb{} ions can substitute Ca$^{2+}$ in two inversion-related  sites, both of them having tetragonal $S_4$ symmetry and magnetically equivalent. For metallic ions in the trivalent oxidation state (such as rare-earth ions in general and \yb{} in particular), charge compensation is necessary to insert them into the \CWO{} lattice in a Ca$^{2+}$ position. Previous studies indicated that the charge compensation mechanism occurs mainly over large distances \cite{Ranon1964_S,Billaud2025_S}, which results in a weak perturbation of the main \yb{} sites, retaining Ca$^{2+}$ $S_4$ site symmetry. This contributes to narrow inhomogeneous broadening observed in our study. However, charge compensation also occurs in the immediate vicinity for a significant part of the \yb{} ions. This mechanism can involve a Ca vacancy or another monovalent ion present in the crystal. This compensation close to the ion can significantly perturb the crystal field at the position of the dopant, modifying its optical and magnetic properties and giving rise to several additional sites appearing in optical and spin resonance measurements.

\section{Sample characterization}
\label{SM:charac}

\subsection{Emission spectra}
\label{SM:opo}

\begin{figure*}[hbtp!]
	\includegraphics[width=0.9\linewidth, trim=0.cm 0.0cm 0.cm 0.0cm,clip]{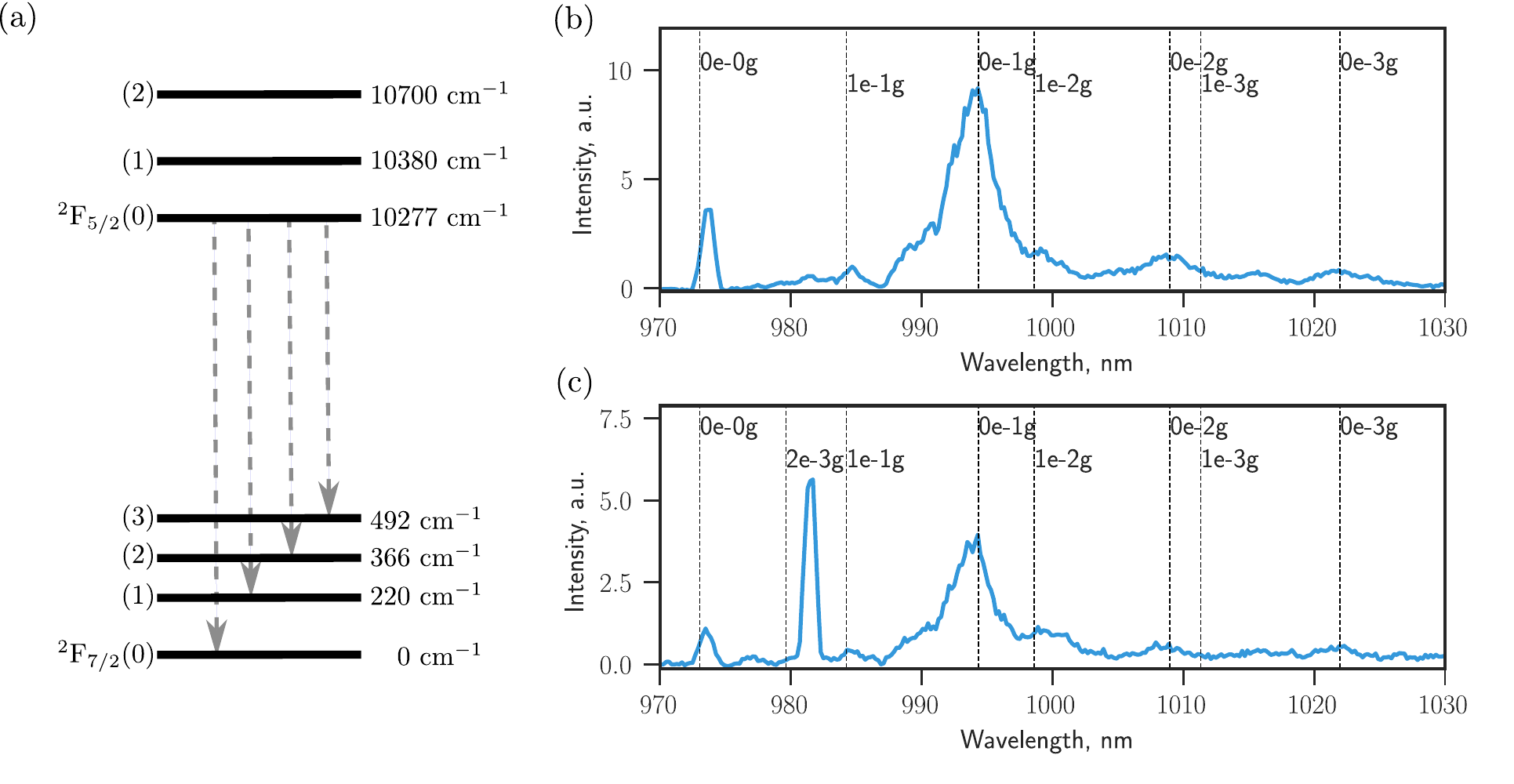}
	\caption{ 
	(color online) \textbf{Crystal field (CF) levels and emission spectra of \ybCWO{}.}
    (a) CF splittings of \yb{} reproduced from \cite{Pappalardo1963_S}.
    Emission spectra of \ybCWO{}  at 10 K excited on the 0 - 1 transition at 963~nm (b) and the 0 - 2 transition at 954~nm (c). Transitions between CF levels are indicated with dashed lines. The line at 981.7 nm is attributed to another \yb{} site.
}
	\label{figSM:opo}
\end{figure*}

The crystal field level structure of \yb{} in \CWO{} was measured previously \cite{Pappalardo1963_S,Jones1967_S} and is indicated in \figref{figSM:opo}(a). We measured the emission spectra of \ybCWO{}  by using a  Nd:YAG pumped OPO laser system with a spectral linewidth  around 1 nm and nanosecond excitation pulse widths at a 10 Hz repetition rate. Excitation using the \dfs(0)$\rightarrow$\dfc(1) transition around 10380 cm$^{-1}$ (963 nm) (\figref{figSM:opo}(b,c)) was performed at a temperature of 15~K. The lines observed corresponding to the \dfc(0)$\rightarrow$\dfs(0,1,2,3) transitions are found according to previous studies. The transition linewidths for the higher energy transitions are much broader than the main transition, which can be due to phonon broadening and vibronic transitions. The emission spectrum, when exciting the \dfs(0)$\rightarrow$\dfc(2) transition around 10700 cm$^{-1}$ (954 nm), contains an additional line around 981.7 nm. The same line is also visible in the absorption spectra shown in the next section. The width of this line is much narrower than the other higher energy lines, which suggests that it comes from another \yb{} site.

\begin{figure*}[hbtp!]
	\includegraphics[width=0.5\linewidth, trim=0.cm 0.0cm 0.cm 0.0cm,clip]{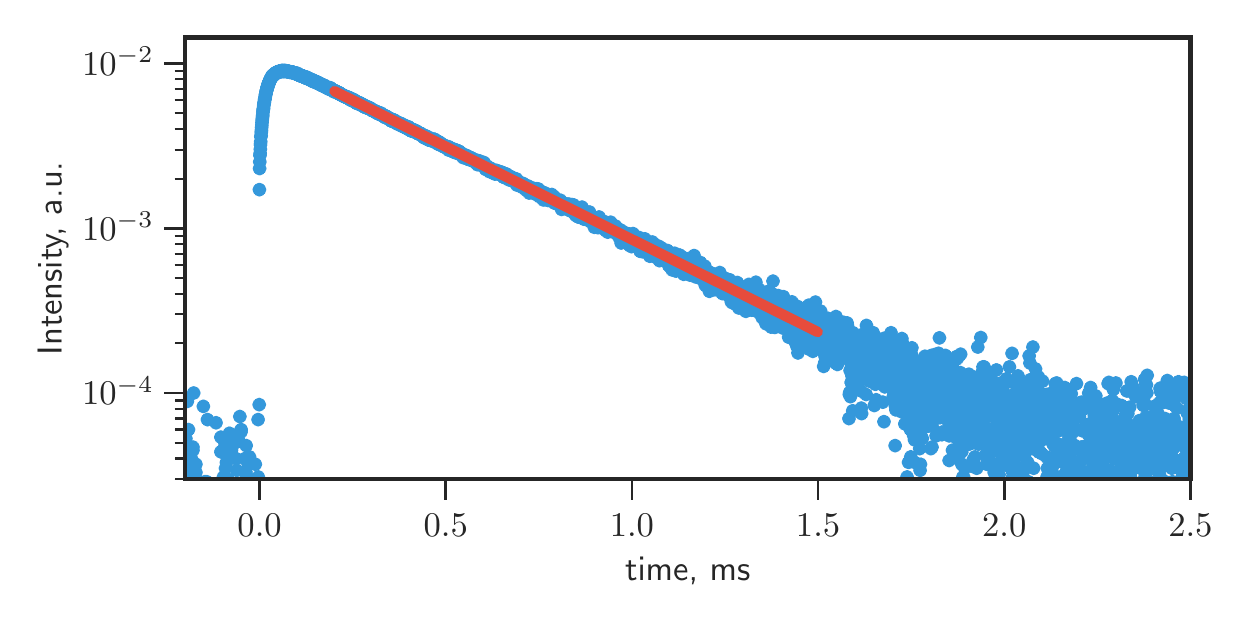}
	\caption{ 
	(color online)  Excited-state \dfc(0) lifetime measurement via fluorescence decay. An exponential fit (solid line) gives 0.385~ms.
}
	\label{figSM:fluo}
\end{figure*}

The excited state lifetime was measured by exciting the \dfs(0)$\rightarrow$\dfc(0) transition and detecting the fluorescence at the \dfc(0)$\rightarrow$\dfs(1) transition wavelength. The result of the exponential fit gives $T_1 = 0.385$~ms (\figref{figSM:fluo}). 

To estimate the branching ratio $\beta$ from the first excited state \dfc(0) to the lowest \dfs(0) ground state level, we integrated the emission spectra from \figref{figSM:opo}(b). The resulting percentage going to the \dfc(0)$\rightarrow$\dfs(0) transition is around 5(1)\%. 
We can also estimate this value by calculating the  spontaneous emission rate $\Gamma_s$ of a two-level emitter
\begin{equation}
\Gamma_s = \frac{2\pi e^2 \nu^2}{\epsilon_0 m_e c^3 } n^2 \chi_L f
\end{equation}
where $\chi_L = (n^2+2)^2/9$, $e$, $m_e$ are the electron charge and mass, $\nu$ is the transition frequency, $n$ the refractive index, $c$ the speed of light,  $\epsilon_0$  the vacuum permittivity, and $f$ is the  oscillator strength of the optical transition. $f$ can be estimated from the absorption spectra using
\begin{equation}
f = \frac{4\pi\epsilon_0 m_e c}{\pi e^2 }\frac{1}{3N} \sum_i{\chi_L^{-1}\alpha_i(\nu)d\nu},
\end{equation}
where $N$ is the ion density and  the summation is done over the three orthogonal polarisations. For \CWO{}, the refractive index along the $a(b)$ axis is $n=1.895$ at 973 nm, and the concentration of \ybiso{} ions in the tetragonal $S_4$ site was estimated using electron paramagnetic resonance measurements to be 4.96 ppm, giving an ion density of $N= 6.97\times 10^{16}$ cm$^{-3}$.
We thus estimated the oscillator strength to be $f = 2.4\times10^{-7}$ giving $\Gamma_s = 60$ s$^{-1}$. This results in the branching ratio being lower bounded by $\beta = \Gamma_s T_1= 0.023$ which corresponds to a purely radiative emission, a reasonable assumption for \yb{} at low concentrations \cite{Welinski2016_S}. The results from optical characterization are summarized in \tabref{SM:tabOpt}.

\subsection{Optical absorption}
\label{SM:cary}

\begin{figure*}[hbt!]
	\includegraphics[width=0.8\linewidth, trim=0.cm 0.0cm 0.cm 0.0cm,clip]{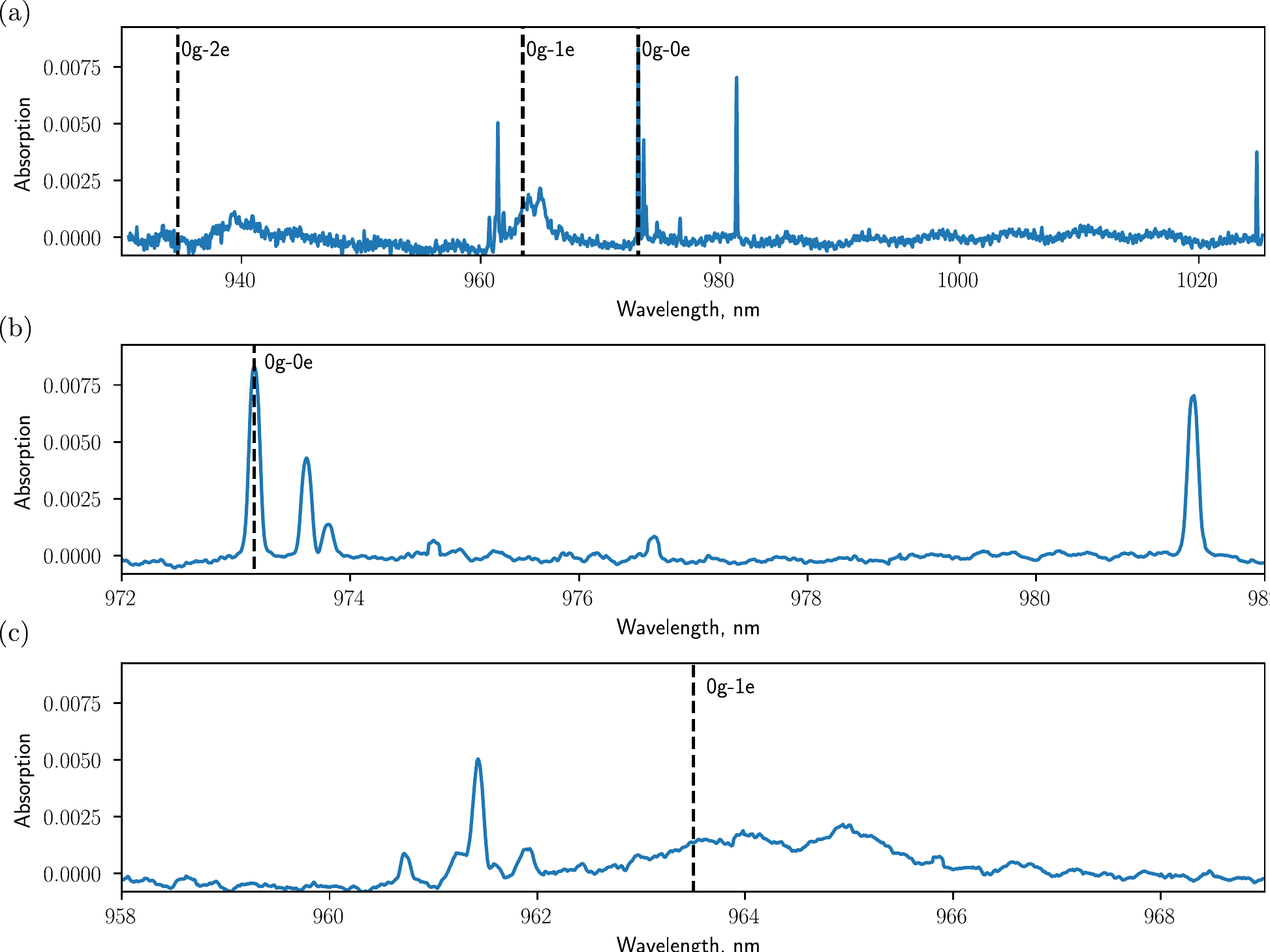}
	\caption{ 
	(color online) \textbf{Absorption spectra of \ybCWO{} at 15K.}
    (a) Absorption spectrum of \ybCWO{} at 15 K with unpolarised light propagating along the $a$ axis. Transitions between different crystal field levels are indicated for the tetragonal $S_4$ site.
    Multiple absorption peaks are visible on the zooms around 977 nm (b) and 963 nm (c).
}
	\label{figSM:cary}
\end{figure*}

The absorption spectrum taken at 15 K is shown in \figref{figSM:cary}.  The position of the  \dfs(0)$\rightarrow$\dfc(0) transition  of the tetragonal $S_4$ site is visible, while absorption lines from the \dfs(0)$\rightarrow$\dfc(1,2) transitions are weaker and slightly shifted with respect to the expected values indicated by dashed lines. The width of the \dfs(0)$\rightarrow$\dfc(0) transition peak is broadened by the resolution of the spectrophotometer (0.1 nm). The other transitions  \dfs(0)$\rightarrow$\dfc(1,2) are homogeneously broadened by phonon relaxation processes with  FWHM between 15 and 30 cm$^{-1}$.

Satellite absorption peaks are visible close to the \dfs(0)$\rightarrow$\dfc(0) transition, and can be attributed to \yb{} sites shifted to lower energy  due to the lower symmetry (\figref{figSM:cary}(b)). They can appear from the charge compensation mechanism in the vicinity of the substitutional site but with different charge configurations leading to multiple lower symmetry orthorhombic $D_2$ sites \cite{Ranon1964_S,Nemarich1968_S}. 
Unidentified absorption peaks around 961.43 nm and 1024.85 nm are also visible in the spectrum in \figref{figSM:cary}(a). The positions of all observed lines are summarised in \tabref{SM:tabOpt}.

\section{Coherence measurements}
\subsection{Experimental setup}
The crystal was placed in a Bluefors dilution cryostat at about $\SI{50}{\milli\kelvin}$. The crystal was fixed on a copper mount using  silver glue, and the mount was screwed to the cold plate of the cryostat cooled down up to 40mK. During the experiment a temperature sensor mounted on the cold plate went up to 60 mK, presumably due to heating through laser scattering.

A $\SI{980}{\nano\metre}$ DLPro Toptica external cavity diode laser was used for high resolution optical spectroscopy and coherence time measurements. We generated optical pulse sequences by controlling  amplitude and frequency using acousto-optical modulators (AOMs). The AOM was driven using the arbitrary waveform generator AWG Keysight M320xA. The laser was split into two paths, namely the pump/probe sent to the crystal and a local oscillator used in a heterodyne detection scheme. The optical power in each path was approximately 5 mW.

For each path, a fiber-based electro-optical phase modulator (iXBLue NIR-MPX-LN-10) was used to add sidebands onto the carrier frequency. The phase modulators (PM) were driven with DS SG4400L and Stanford Research Systems SG390 rf generators, amplified to the 1 W power level. A series of TTL switches was used to alternate between several frequencies generating sidebands using the PM. The TTL switches and all the devices were synchronized using a SpinCore PulseBlaster module. 

The pulse sequence consisted of an optical pumping step (300 ms), a delay (10 ms), and the coherence measurement (up to 600 ms). The entire sequence was run at a 1 Hz repetition rate.

\begin{figure*}[hbtp!]
	\includegraphics[width=0.65\linewidth, trim=0.cm 0.03cm 0.cm 0.0cm,clip]{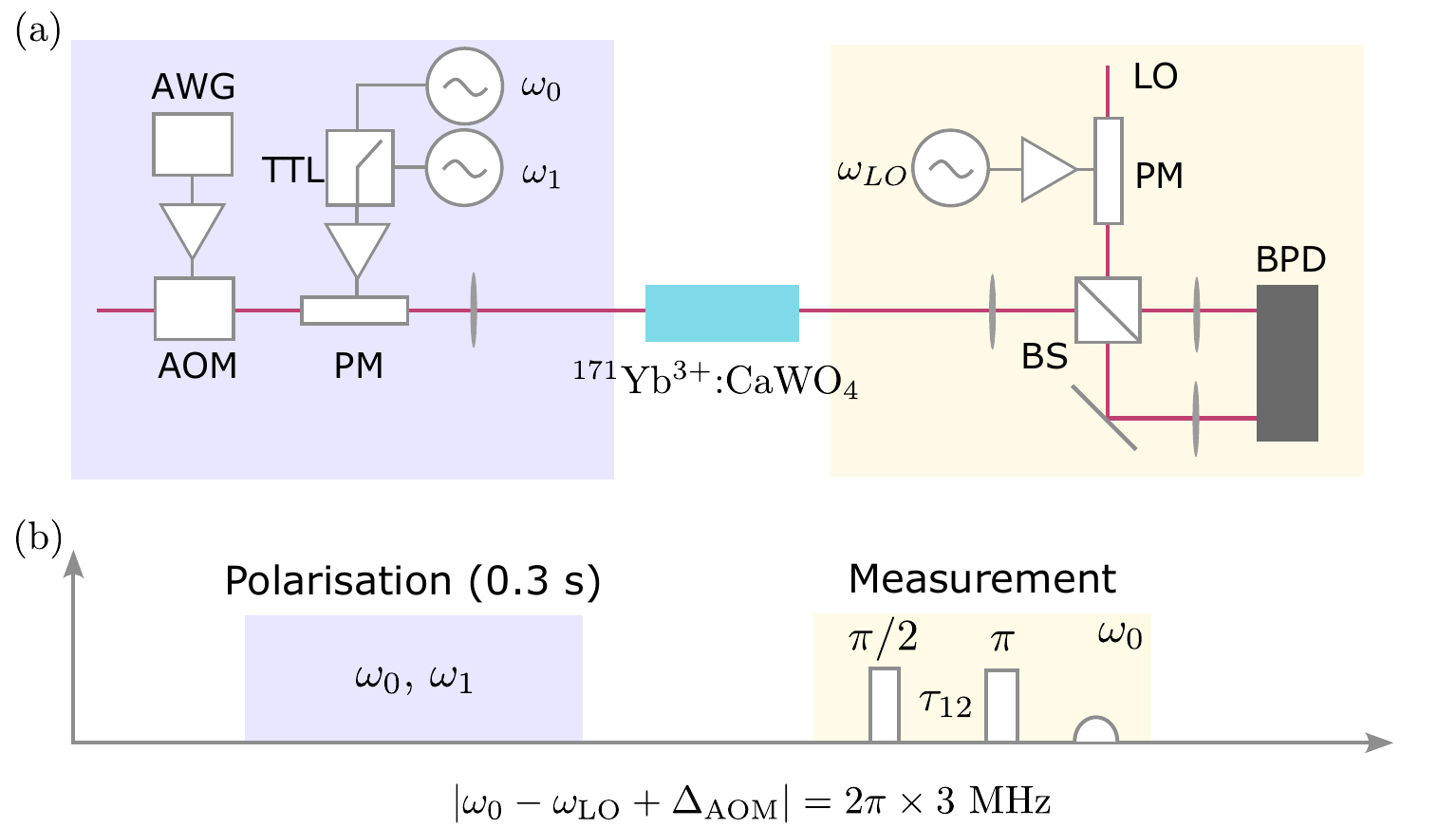}
	\caption{ 
	(color online) Experimental setup (a) and time sequence of the experiment (see text for details). AWG - arbitrary waveform generator, AOM - acousto-optic modulator, PM - phase modulator, TTL- microwave switch, BPD - balanced phot diode, BS -  beamsplitter.
}
	\label{figSM:expsetup}
\end{figure*}

The laser beam was focused into the crystal using a free-space path through the cryostat. A lens with 30 cm focal length was used to focus the laser beam down to a 0.2 mm diameter. The local oscillator was interfered with the optical probe after the cryostat. The beam was later detected using a Thorlabs PDB450C balanced photodiode. The detection is performed at 3~MHz given by  the difference between two microwave generators driving PMs (\figref{figSM:expsetup}).

\subsection{High-resolution optical absorption}
\label{SM:abs}

\begin{figure*}[hbtp!]
	\includegraphics[width=0.99\linewidth, trim=0.cm 0.03cm 0.cm 0.0cm,clip]{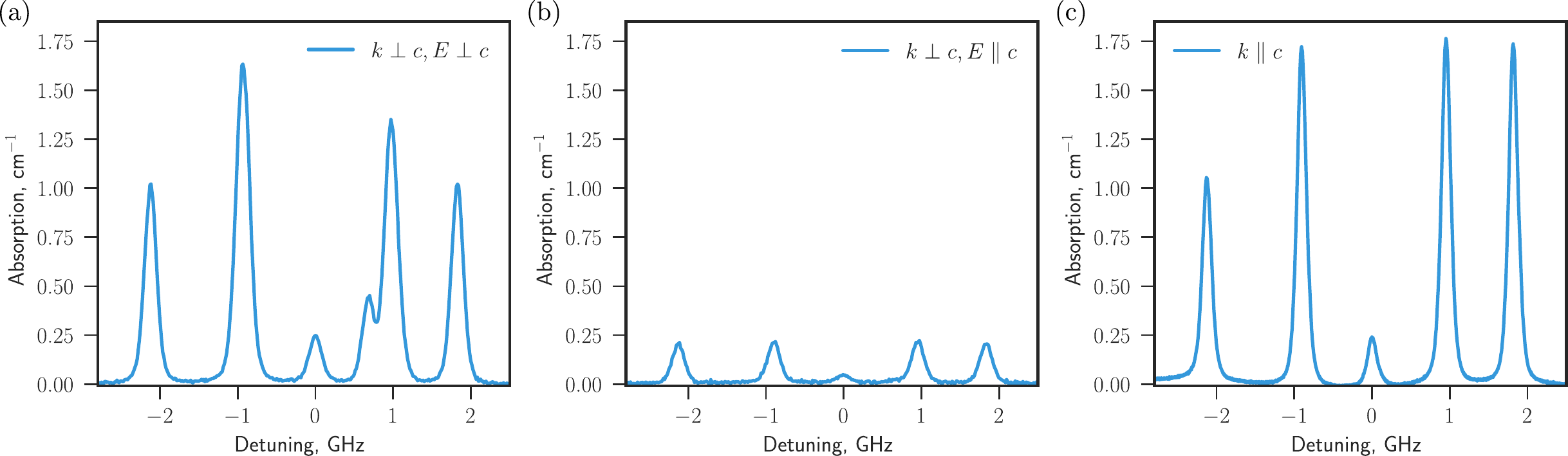}
	\caption{ 
	(color online) High-resolution optical absorption spectra of the \transition{} transition for the tetragonal site of \ybCWO{} recorded for $k \perp c$ and light polarised  perpendicular  (a) and parallel (b) to the $c$ axis, as well as a spectrum taken with $k \parallel c$ (c). The absorption spectra in (a) and (b) were measured at 4.6~K, while (c) was measured at 50~mK cryogenic temperatures.
}
	\label{figSM:abs}
\end{figure*}

High-resolution absorption spectra for the \transition{} transitions of the tetragonal $S_4$ site were measured at zero magnetic fields for various polarisation orientations (\figref{figSM:abs}). For this, the laser was scanned using the piezo actuator. Absorption scans were taken for the electric field $E$ perpendicular to $c$ axis ($E \perp c$),  along the $c$ axis  ($E \parallel c$), and with the wave vector $k$ parallel to $c$ ($k \parallel c$) (\figref{figSM:abs}(a-c)).

The spectra consist of multiple narrow absorption peaks corresponding to transitions connecting various hyperfine levels in the ground and excited state for \ybiso{} ions. The central peak of the spectra at zero detuning contains the lines corresponding to $I=0$ \yb{} isotopes. The ratio between the total absorption of \ybiso{}  $I=1/2$ isotope and $I=0$ isotopes gives around 5 \% abundance of  $I=0$ isotopes in accordance with the purity of the Yb$_2$O$_3$ powder used during the crystal growth. The $^{173}$\yb{} isotope  ($I=5/2$)  was not visible in the optical absorption spectra. All lines could be well fit by a Gaussian linewidth with a FWHM of 185~MHz. Such narrow inhomogeneous broadening, much lower than the hyperfine energy splittings, allows  resolving the hyperfine structure in the ground and excited states and separately addressing different spin levels. This results in the possibility  to optically pump the whole spin ensemble (see Section 4.D).

Due to the non-polar site symmetry of \CWO{}, the linear electric field shifts of the optical transitions are suppressed. This reduces the effect of electric field variations along the crystal, that can partly explain  inhomogeneous broadening down to 140 MHz for the $k \parallel c$ absorption spectra (\figref{figSM:abs}(c)). This measurement was done using the same sample but during a different cool down of the cryostat and after regluing the sample to the cold plate. The observed $\Gamma_{inh}$ variation can be attributed to the strain inside  crystal samples. 

\setlength{\tabcolsep}{0.1em} 
\renewcommand{\arraystretch}{1.25}
\begin{table*}
 \caption{\label{optTable} Optical transition parameters for the tetragonal site in \ybCWO{}. The listed parameters include: CF level energies, \transition{} transition wavelength (vac.) ($\lambda_{vac}$), inhomogeneous linewidths ($\Gamma_{inh}$), the integrated peak absorption coefficient for different light polarisations ($\alpha_0$), oscillator strength ($f$), experimental fluorescence decay time ($T_1$), and the branching ratio for the 0-0 transition ($\beta$). Optical transitions for three orthorhombic sites of \ybCWO{} are given, together with unidentified lines appearing in the absorption spectra at 10 K (\figref{figSM:cary}).
 }
\begin{ruledtabular}
 \begin{tabular}{ccccccccc}
         & \multicolumn{2}{c}{Energy, cm$^{-1}$} & $\lambda_{vac}$, nm & $\Gamma_\mathrm{inh}$, GHz & $\alpha_0$, cm$^{-1}$ & $f \times 10^7$ & $T_1$, ms & $\beta$ \\
& \dfs & \dfc \\        
$S_4$ site & 0 & 10277 & 973.162 & 0.185 ($E \perp c$) &  5.3 ($E \perp c$)  & 2.4 & 0.385 &  0.023 \\
        & 220 & 10380 &&  0.185 ($E \parallel c$) & 0.78 ($E \parallel c$) & \\
        & 366 &  10700 &&  0.140 ($k \parallel c$) & 5.95 ($k \parallel c$) & \\
        & 492 \\
        \\
  Other sites    & & 10270.98 & 973.62 \\
                  & & 10268.95 & 973.81 \\
                 & & 10189.83  & 981.37  \\
  Unidentified    & & 10401.2 & 961.43 \\
                  & & 9757.52 & 1024.85 \\
                 
 \end{tabular}
 \label{SM:tabOpt}
\end{ruledtabular}
 \end{table*}

\subsection{Magnetic field sweeps.}
The bulk crystals were mounted in an u-bench on the cold plate of a dilution refrigerator. The base temperature during these measurements was $\approx800$ mK. We mounted a set of home-built Helmholtz coils either parallel or perpendicular to the crystalline c-axis and  applied a current up to $\pm 10$ A in steps of $0.1$ A using a magnet power supply system (American Magnetics Inc.). Light from a tunable diode laser (Toptica CTL) was coupled to a fiber and sent through a set of polarisation paddles before being directed into the u-bench setup. 

To sweep the laser frequency over the entire absorption spectrum, we used a triangle-shaped voltage waveform to modulate the piezo actuator of the laser. The piezo modulation and the transmission through the crystal were recorded on an oscilloscope, which allowed us to extract the absorption as a function of piezo voltage. To relate this to the frequency of the laser, a copy of the laser light was sent through a fiber Fabry-Perot cavity and its transmission recorded. The fiber cavity had a calibrated free spectral range of 136 MHz.  By interpolating the frequency between the transmission peaks of the fiber cavity,  a relative frequency axis for each magnetic field setting was obtained. As the laser could drift over the time scale of the experiment, the center frequency of the laser was measured during the sweep using a wavemeter (Bristol 771) in order to obtain an absolute frequency axis.

The eigenstates of the ground and excited states were described by the following Hamiltonian: $ H_{(g/e)} = \mathbf{I} \cdot \mathbf{A} \cdot \mathbf{S} + \mu_B \mathbf{B} \cdot \mathbf{g_{g/e}} \cdot \mathbf{S}$, ignoring the nuclear Zeeman effect ($\mu_B \gg \mu_n$). We calculated the frequencies of all transitions between the ground and excited states and convoluted them with a Gaussian distribution with a full-width at half-maximum of 136 MHz for the $^{171}$Yb isotope and 153 MHz for the zero nuclear spin (ZNS) isotopes. Both values were obtained from fitting absorption scans. As a perfectly aligned magnetic field was assumed, the only free parameters in the fit are the excited state g-tensor, a scaling factor to relate the magnet current to the magnetic field, the amplitude of the absorption dips associated with \ybiso{} and the zero nuclear spin isotopes, and an overall frequency offset. We performed the fit using a least-squares method (lmfit python package).

\begin{figure*}[hbtp!]
	\includegraphics[width=0.79\linewidth, trim=0.cm 0.0cm 0.cm 0.0cm,clip]{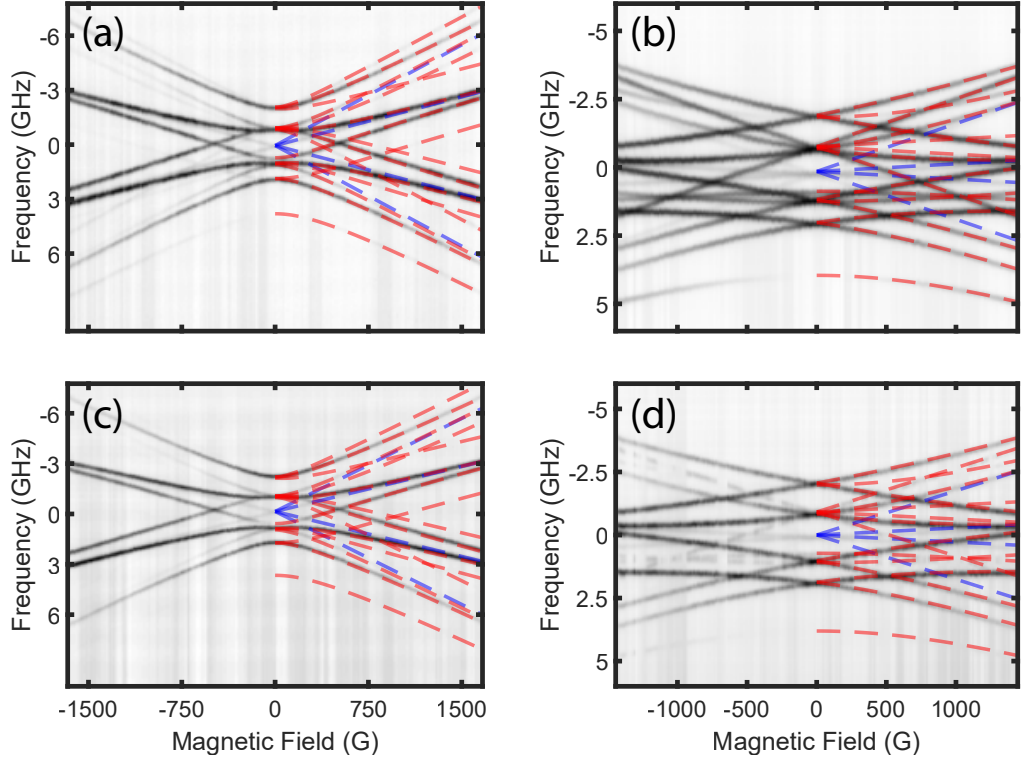}
	\caption{ 
	(color online) Magnetic field dependence of the optical transitions between hyperfine levels  in \ybCWO{}. Absorption spectra for varying magnetic field strengths applied parallel to the crystalline $a$-axis (left column) and parallel to the  $a$-axis (right column). The upper and lower rows show scans with $E \perp c$, respectively $E \parallel c$. Dashed lines show fits to a model given by a spin Hamiltonian (see text). Red and blue lines correspond to  \ybiso{} and  $I=0$ isotopes. 
}
	\label{figSM:Bsweep}
\end{figure*}

\begin{figure*}[hbtp!]
	\includegraphics[width=0.79\linewidth, trim=0.cm 0.0cm 0.cm 0.0cm,clip]{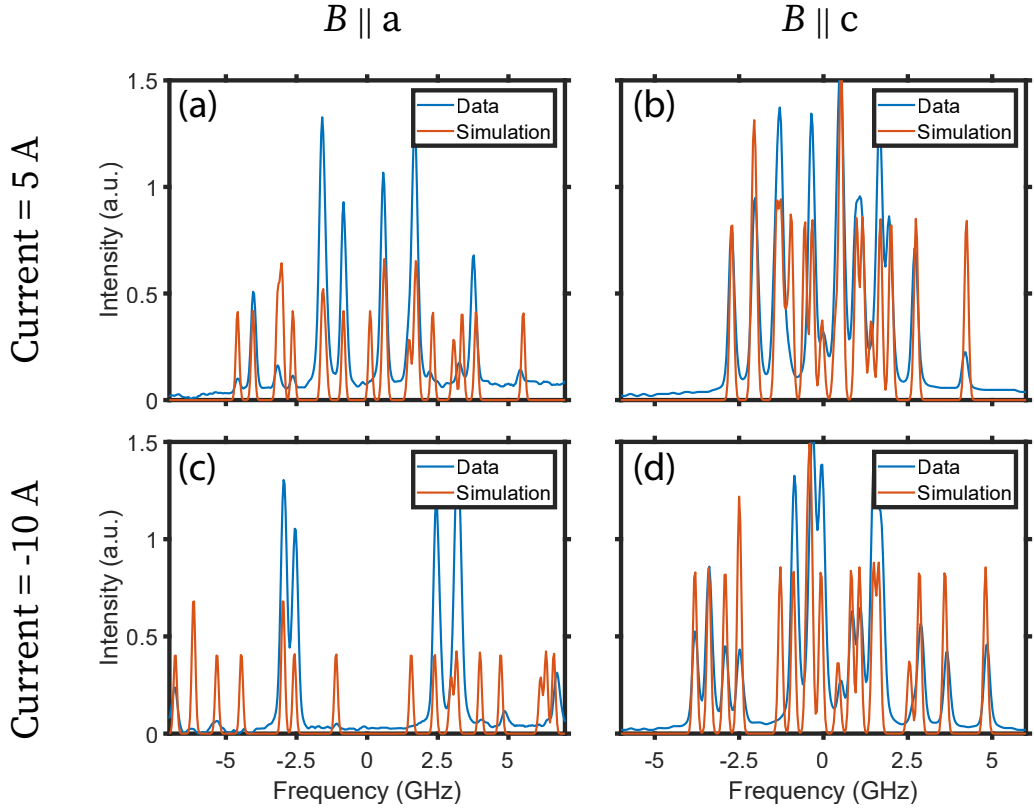}
	\caption{ 
	(color online) Absorption spectra for fixed magnetic fields applied perpendicular or parallel to the crystalline c-axis (left and right columns, blue lines). The top row shows spectra acquired under fields of 829 G (left) and 718 G (right), whereas the bottom row shows spectra recorded under fields of  1659 G (left) and 1436 G (right). The simulated spectra (red lines) assume all transitions are of equal strength.  
}
	\label{figSM:Bsweep2}
\end{figure*}

The extracted g-tensor principal values and magnetic field scaling for light polarised perpendicular or parallel to $c$ are listed in \tabref{tabSM:Bsweep}. As expected, the values are very close in both cases. \figref{figSM:Bsweep} shows the acquired data overlayed with the fitted curves, for both polarisations and magnetic field directions. Absorption scans for two values of the magnetic field are plotted in \figref{figSM:Bsweep2}, together with the numerical model assuming all optical transitions to be allowed with equal strength. 

\setlength{\tabcolsep}{0.1em} 
\renewcommand{\arraystretch}{1.25}
\begin{table*}
 \caption{ Extracted parameters from the fit of the magnetic field dataset. Scaling factor $s$ relates the current through to coils to the applied magnetic field. The datasets acquired for the different polarisations ($E \parallel c$ and $E \perp c$) should yield the same results.
 }
\begin{ruledtabular}
\begin{minipage}[t]{.5\linewidth}
 \begin{tabular}{ccc}
   & 	$E \parallel c$	& $E \perp c$  \\  \hline
$g_{e,\parallel}$   & -1.453(1) & -1.451(1) \\
$s_{\parallel}$   & 143.63(8) G/A & 143.64(5) G/A \\
$g_{e,\perp}$   & 1.363(1) & 1.361(1)  \\
$s_{\perp}$   & 165.52(2) G/A & 166.20(2) G/A \\               
 \end{tabular}
 \label{tabSM:Bsweep}
 \end{minipage}
\end{ruledtabular}
 \end{table*}

\setlength{\tabcolsep}{0.4em} 
\renewcommand{\arraystretch}{1.25}
\begin{table*}[!htb]
\caption{Relative branching ratios for optical transitions between hyperfine levels for the \ybCWO{} \transition{} transition of the tetragonal $S_4$ site. Measurements were performed for light polarised perpendicular to the $c$-axis (left) and parallel to the $c$ axis (center), and with $k \parallel c$ (right). 
}
\begin{ruledtabular}
\begin{minipage}[t]{.325\linewidth}
    \begin{tabular}{c|ccc}
            &  $\ket{1,2}_e$ & $\ket{3}_e$ & $\ket{4}_e$  \\ 
            \cline{1-4}
            {$\bra{1}_g$} & 0.3 &  0.7 & 0.0  \\
            {$\bra{2,3}_g$} & 1 & 0.3 & 0.7 \\
            {$\bra{4}_g$} & 0.7 & 0.0 & 0.3 \\
        \end{tabular}
\end{minipage}%
\end{ruledtabular}
\begin{ruledtabular}
\begin{minipage}[t]{.325\linewidth}
    \begin{tabular}{c|ccc}
            &  $\ket{1,2}_e$ & $\ket{3}_e$ & $\ket{4}_e$  \\ 
            \cline{1-4}
            {$\bra{1}_g$} & 1.0 &  0.0 & 0.0  \\
            {$\bra{2,3}_g$} & 0 & 1.0 & 1.0 \\
            {$\bra{4}_g$} & 1.0 & 0.0 & 0.0 \\
        \end{tabular}
\end{minipage}%
\end{ruledtabular}
\begin{ruledtabular}
\begin{minipage}[t]{.325\linewidth}
    \begin{tabular}{c|ccc}
            &  $\ket{1,2}_e$ & $\ket{3}_e$ & $\ket{4}_e$  \\ 
            \cline{1-4}
            {$\bra{1}_g$} & 1.0 &  0.0 & 0.0  \\
            {$\bra{2,3}_g$} & 0 & 1.0 & 1.0 \\
            {$\bra{4}_g$} & 1.0 & 0.0 & 0.0 \\
        \end{tabular}
\end{minipage}%
\end{ruledtabular}
\label{SM:optTable}
\end{table*}

\subsection{Optical pumping}
\label{SM:pump}

\begin{figure*}[hbtp!]
	\includegraphics[width=0.99\linewidth, trim=0.cm 0.0cm 0.cm 0.0cm,clip]{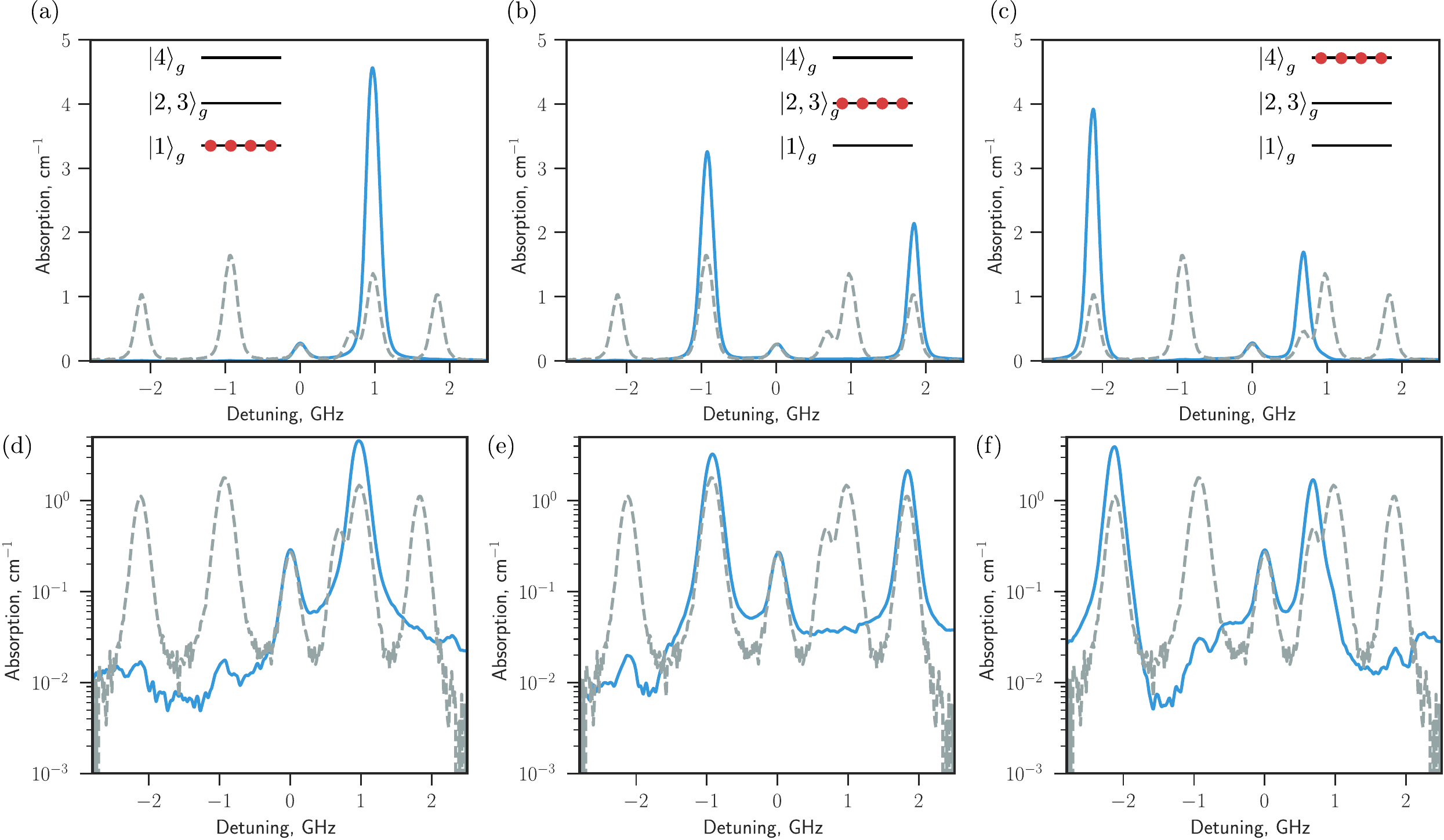}
	\caption{ 
	(color online) High-resolution optical absorption spectra of the \transition{} transition for tetragonal site of \ybCWO{} recorded for light polarised perpendicular to $c$ axis pumping ions into $\ket{1}_g$ (a,d), $\ket{2,3}_g$ (b,e), and $\ket{4}_g$ (c,f) states. Absorption spectra without repumping are shown with a dashed gray line.
}
	\label{figSM:repump}
\end{figure*}

To extract the ground and excited state energy splittings and the relative intensities of each transition, we performed an optical pumping experiment at 50 mK temperature. An intense 0.3 s optical pulse was sent to the crystal using a laser resonant with one of the absorption peaks. The absorption profile after pumping into different initial states was then recorded after scanning the laser across all lines (\figref{figSM:repump}). 

We extracted the energy splittings in the excited state by analyzing the relative positions of all the lines. We verified the order of the energy levels in the ground state, for which the hyperfine $A$ tensor was previously measured by \cite{Ranon1964_S, Rakhmatullin2009_S}.
The absorption peak after pumping into the $\ket{1}_g$ state \figref{figSM:repump}(a) contains two transitions with a splitting around 0.08 GHz, resulting in an overall broader inhomogeneous linewidth. The same is true for the peak on the left in  \figref{figSM:repump}(b), when pumping into $\ket{2,3}_g$ states. For the left-most absorption peak in \figref{figSM:repump}(c) the linewidth is as narrow as for the other isolated peaks and equal to 0.185 GHz, suggesting that only one optical transition contributes to it.

The absorption peak at the zero laser detuning was not affected by the optical pumping sequences. It was the same for all the measured absorption spectra, confirming that it comes from \yb{} isotopes with $I=0$.

The optical initialization efficiency into $\ket{1}_g$ state  was estimated to be $>99$\% after integrating the absorption corresponding to $\ket{2,3}_g$ and $\ket{4}_g$  states \figref{figSM:repump}(d).

By fitting all the lines with fixed FWHM linewidths of 0.185 GHz, we extracted the relative transition intensities for transitions connecting a given ground state to different excited states. The result of doing it for all ground states is given in the \tabref{SM:optTable}. Some optical transitions in the hyperfine spectra are forbidden, leading to almost diagonal branching ratio tables for $\pi$ ($E \parallel c$) and $\alpha$ ($k \parallel c$) polarisations. For $\sigma$ ($E \perp c$) polarisation, additional lines appear, potentially coming from magnetic dipole optical transitions (see Section 6.D).

\subsection{Spin-lattice relaxation (SLR) at zero magnetic field}
\label{SM:SLR}

The ability to polarise the whole spin ensemble via optical pumping allowed us to characterize the spin relaxation process through the spin-lattice interaction at temperatures down to 50~mK. For this, we initialized the spin ensemble in the $\ket{1}_g$ state,  the same way it was done in the previous section. Polarisation above 99\% was achieved at the lowest temperatures. We repeated the absorption spectra measurement after variable time delay after the polarisation (from millisecond to a few hours) to extract the spin dynamics due to the spin-lattice interaction at zero magnetic field and variable temperatures between 50 mK and 3.6 K. The absorption spectra were analyzed for each delay to extract populations $n_{1g}$, $n_{2,3g}$, and $n_{4g}$ in $\ket{1}_g$, $\ket{2,3}_g$, and $\ket{4}_g$, respectively. The recovery of the perturbed spin population was measured towards the temperature equilibrium distribution of the spin ensemble $n^{\text{eq}}_{ig}$ given by the energy spacings between different levels  and the equilibrium temperature $T^{\text{eq}}$ by $\exp{-E_i/(k_BT^{\text{eq}})}$, where $k_B$ is the Boltzmann constant and $E_i$ is the energy for state $i$.

\begin{figure*}[hbtp!]
	\includegraphics[width=0.7\linewidth, trim=0.cm 0.03cm 0.cm 0.0cm,clip]{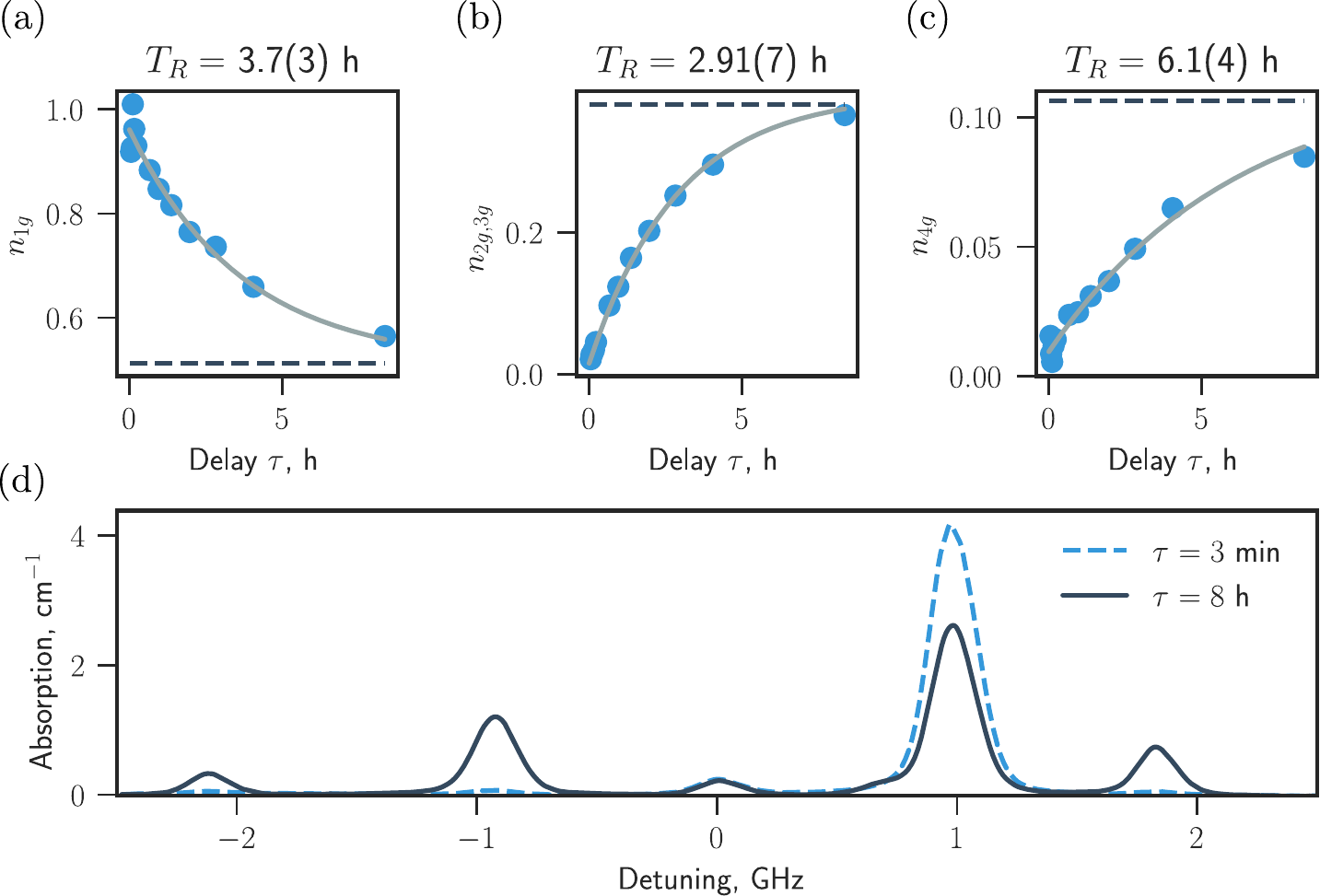}
	\caption{ 
	(color online) \textbf{Spin-lattice relaxation measurement at 50 mK temperature}. Spin populations in different levels $n_{1g}$ (a), $n_{2,3g}$ (b), and $n_{4g}$ (c) measured for different delay times $\tau$ after the spin polarisation step. The equilibrium temperature of the spin ensemble of $T^{\text{eq}}=140$ mK was extracted from the fitting. The dashed lines indicate the population equilibrium values at $T^{\text{eq}}$. (d) An example absorption spectra measured after the first delay ($\tau=3$ minutes) and last delay  ($\tau=8$ hours).
}
	\label{figSM:slrDecay}
\end{figure*}

\begin{figure*}[hbtp!]
	\includegraphics[width=0.7\linewidth, trim=0.cm 0.0cm 0.cm 0.0cm,clip]{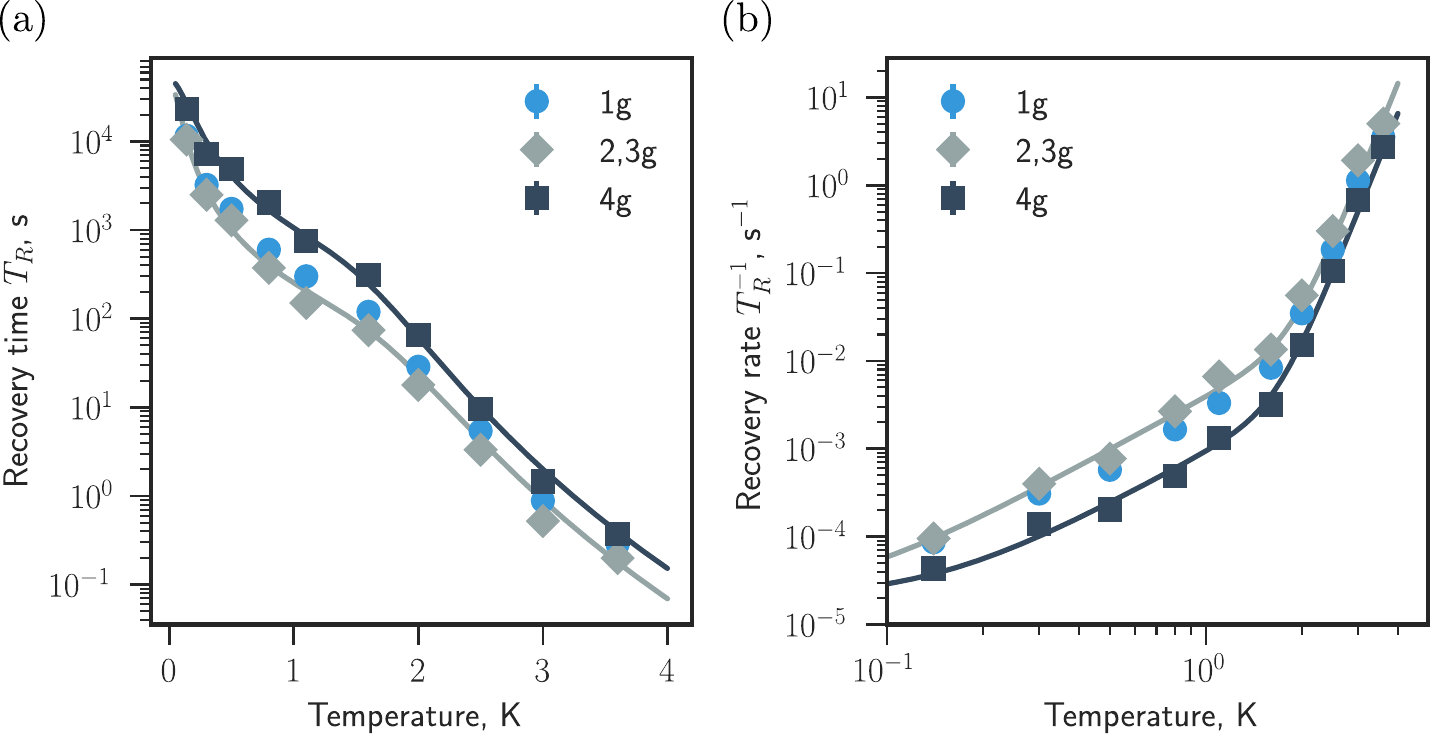}
	\caption{ 
	(color online) \textbf{Spin-lattice relaxation at different temperatures}. The extracted spin population recovery times $T_R$ (a) and recovery rates $T_R^{-1}$ (b)  for different levels measured at different temperatures. The temperature axis represents the equilibrium temperatures $T^{\text{eq}}$  extracted from the fit. The solid lines are fits for $\ket{2,3}_g$ and $\ket{4}_g$ population using the model from the text.
}
	\label{figSM:slrTemp}
\end{figure*}

The spin population recovery at the lowest temperature of the cryostat (50 mK) is shown on \figref{figSM:slrDecay}. The recovery times are extracted using exponential fits. The equilibrium populations in the fits were shared for populations from different levels. The extracted population distribution corresponds to an equilibrium temperature of $T^{\text{eq}} = 140$ mK. A higher value for $T^{\text{eq}}$ compared to the 50 mK cryostat temperature can be due to the additional heat load from the optical windows of the cryostat and/or to phonon bottleneck effects in the crystal \cite{Kukharchyk2020_S}.

The temperature dependence of the recovery time $T_R$ was measured up to 3.6K temperature \figref{figSM:slrTemp}. A higher relaxation rate is observed at higher temperatures due to the spin-lattice relaxation mechanism. Below 1.5~K, the relaxation decreases quadratically with the temperature and phonon relaxation is expected to be limited by the direct phonon relaxation process induced by the hyperfine interaction \cite{Baker1964_S}. For temperatures above 2 K, the relaxation is due to the two-phonon Raman process \cite{Abragam1970_S}(Chapter 10) \cite{Chiossi2024_S}. The temperature dependence of the recovery rate was fitted using $R_{\text{ph}} = R_0 + a_1T^2 + a_2T^9$, where the first term can be related to residual excitation from the environment, second term represents the direct phonon process limited by phonon bottleneck effect, and the third is given by the two-phonon Raman relaxation process. The best-fit results are given in \tabref{SM:tabPhonon}.

\setlength{\tabcolsep}{0.1em} 
\renewcommand{\arraystretch}{1.25}
\begin{table}
 \caption{Spin lattice relaxation (SLR) parameters for \ybCWO{}. $a_1$ and $a_2$ correspond to the direct and Raman processes.
 }
\begin{ruledtabular}
\begin{minipage}[t]{.5\linewidth}
 \begin{tabular}{c|ccc}
     & $R_0 \times 10^{-4}$ Hz & $a_1 \times 10^{-4}$Hz K$^{-2}$ & $a_2 \times 10^{-4}$ Hz K$^{-9}$  \\ \hline
 $n_{2,3g}$    & 0.2   &  3.8   &  0.55  \\
 $n_{4g}$    &  0.2   &  9   &  0.25  \\
 \end{tabular}
 \label{SM:tabPhonon}
 \end{minipage}
\end{ruledtabular}
 \end{table}

\subsection{Optically detected spin echo}
We measured the spin coherence time $T_2^\text{s}$ through optical detection of a spin echo in a Hahn echo sequence using all-optical spin manipulation via Raman transitions through the excited state \cite{Rancic2017_S}. All spin echo measurements reported in this paper were carried out on the $\ket{1}_{g}\rightarrow\ket{4}_{g}$ transition (${3083.85}$ MHz) of the tetragonal $S_4$ site. Without compensating the Earth's magnetic field, a spin coherence time of $30$ ms was measured. A permanent magnet was  used to compensate the external magnetic field and optimise the coherence properties by maximising the spin echo signal at long delays.

The duration of optical pulses in the spin echo sequence was $\SI{2}{\micro\second}$ and was optimised using the spin echo intensity. All pulses had a Gaussian envelope shape. Optical pulses of $\SI{200}{\micro\second}$ duration were used to resolve the spin linewidth, which was found to be  $5(1)$~kHz FWHM. 
The detected signal was analysed with a Fast Fourier Transform and the area of the peak corresponding to the beat at $|\omega_0 - \omega_{LO} + \Delta_{\text{AOM}}|$ (\figref{figSM:expsetup}) was acquired at various time delays $\tau_{12}$. The coherence time $T_2^\text{s}$ was then extracted by fitting the peak area decays as $E_\text{echo}(\tau)=E_0 \exp(-2\tau_{12}/T_2^\text{s})$.

\subsection{Photon echo}

The optical coherence time $T_2^{\text{o}}$ was measured in an analogous way by preparing a photon echo sequence on a transition between a ground and an excited state. We used the heterodyne signal to measure the echo intensity detected at $|\nu_\text{LO}-\nu_0|$ to improve the detection sensitivity. The duration of the optical pulses was about $\SI{1}{\micro\second}$. The optical coherence time $T_2^\text{o}$ was extracted by fitting photon echo area using $E_\text{echo}(\tau)=E_0 \exp(-2\tau_{12}/T_2^\text{o})$. 

\section{Spin Hamiltonian}
\label{SM:Hs}

The optical transition \transition{} connects the  lowest energy crystal field doublets of the ground state and the excited states. These doublets can be described using an effective  $S = 1/2$ spin Hamiltonian involving the interaction with an external magnetic field $\mathbf{B}$~\cite{Abragam1970_S} (Chapter 1). In this work, we focus on the ${}^{171}\mathrm{Yb}$ isotope, which has a nuclear spin $I = 1/2$, such that the effective spin Hamiltonian for the ground and excited states can be written as 
\begin{equation}
\label{eq:Heff1}
\mathcal{H} = \mathbf{I} \cdot \mathbf{A} \cdot \mathbf{S} + \mu_\text{B} \mathbf{B} \cdot \mathbf{g} \cdot \mathbf{S} - \mu_\text{n}  \mathbf{B}\cdot \mathbf{g}_\text{n} \cdot \mathbf{I},
\end{equation}
where $\mathbf{g}$ and $\mathbf{A}$ are the coupling tensors of the electronic Zeeman and hyperfine interactions, respectively, $\mu_B$ and $\mu_n$ are the electronic and nuclear spin magnetons. The nuclear Zeeman interaction $\mathbf{g}_\text{n}$ tensor is considered to be isotropic with $g_n= 0.987$ for \ybiso{} nuclear spin.

The first term is due to the hyperfine coupling between the electron and nuclear spin, where $\mathbf{I}$ is the nuclear spin operator. The second term describes the electronic Zeeman interaction, where $\mu_B$ is the Bohr magneton, and $\bf{B}$ is the applied magnetic field. The last term arises from the nuclear Zeeman interaction, where $\mu_n$ is the nuclear magneton and $\bf{g_n}$ is the nuclear Zeeman tensor.

Due to the symmetry of the crystal field for the tetragonal $S_4$ site, the spin Hamiltonian can be simplified to
\begin{align}
\label{eq:Heff2}
\mathcal{H} = A_{\perp} (S_x I_x+ S_y I_y)+A_{\parallel} S_z I_z + \mu_\text{B}(g_{\perp} (B_x S_x+ B_y S_y)+g_{\parallel} B_z S_z) - \\
- \mu_\text{n}g_n(B_x I_x+ B_y I_y + B_z I_z),
\end{align}
where $A_{\perp}$ and $g_{\perp}$ ($A_{\parallel}$ and $g_{\parallel}$) are the  components of the $\mathbf{A}$ and $\mathbf{g}$ tensors  corresponding to  directions perpendicular (parallel) to the $c$ axis of the crystal, aligned with the $z$ axis in the above expression.

At zero magnetic field, $\mathbf{B}=0$, the zero field energy splittings are given by the hyperfine interaction $\mathbf{I} \cdot \mathbf{A} \cdot \mathbf{S}$ eigenvalues:  
\begin{align}
&E_{1g} = (-A_{\parallel g} - 2A_{\perp g})/4, &  &E_{1e,2e} = A_{\parallel e}/4,\\
&E_{2g,3g} = A_{\parallel g}/4, & &E_{3e} = (-A_{\parallel e} - 2A_{\perp e})/4, \\
&E_{4g} = (-A_{\parallel g} + 2A_{\perp g})/4, & &E_{4e} = (-A_{\parallel e} + 2A_{\perp e})/4.
\end{align}
Denoting the electron spin  states as $\ket{\uparrow} = \ket{S_z=+1/2}$, $\ket{\downarrow} = \ket{S_z=-1/2}$ and nuclear spin components as $\ket{\Uparrow} = \ket{I_z=+1/2}$, $\ket{\Downarrow} = \ket{I_z=-1/2}$, the corresponding eigenstates at zero magnetic field can be written as
\begin{align}
&\ket{1}_g = (\ket{\uparrow \Downarrow }_g - \ket{\downarrow \Uparrow }_g)/\sqrt{2}, & &\ket{1,2}_e = \ket{\uparrow \Uparrow }_e, \ket{\downarrow \Downarrow }_e, \\
&\ket{2,3}_g = \ket{\uparrow \Uparrow }_g, \ket{\downarrow \Downarrow }_g, & &\ket{3}_e = (\ket{\uparrow \Downarrow }_e - \ket{ \downarrow \Uparrow}_e)/\sqrt{2}, &\\
&\ket{4}_g = (\ket{\uparrow \Downarrow  }_g + \ket{\downarrow \Uparrow }_g)/\sqrt{2}, & &\ket{4}_e = (\ket{\uparrow \Downarrow }_e + \ket{\downarrow \Uparrow }_e)/\sqrt{2}.
\end{align}
Applying a strong magnetic field along the $c$-axis will purify the eigenstates through the Zeeman interaction such that
\begin{align}
&\ket{1}_g = \ket{\downarrow \Uparrow }_g & &\ket{1}_e = \ket{\downarrow \Downarrow }_e, \\
&\ket{2}_g = \ket{\downarrow \Downarrow }_g, & &\ket{2}_e = \ket{\downarrow \Uparrow }_e, \\
&\ket{3}_g = \ket{\uparrow \Downarrow }_g, & &\ket{3}_e = \ket{\uparrow \Uparrow }_e, \\
&\ket{4}_g = \ket{\uparrow \Uparrow }_g, & &\ket{4}_e = \ket{\uparrow \Downarrow }_e.
\end{align}
The states are numbered from lowest to highest energies for the ground and excited state manifolds.

Due to the high symmetry of the crystal field, the degeneracy at zero magnetic field is not completely removed, leading to Zeeman doublets  in both the ground and excited states. The magnetic field sensitivity of the levels' energies, in this case, strongly depends on the electron-nuclear wavefunction. The first order sensitivity for state $\ket{i}$ can be estimated using $\bra{i} \mu_\text{B} \mathbf{B} \cdot \mathbf{g} \cdot \mathbf{S} - \mu_\text{n}  \mathbf{B}\cdot \mathbf{g}_\text{n} \cdot \mathbf{I} \ket{i}$.
Non-degenerate states $\ket{1}_g$,  $\ket{4}_g$, and $\ket{3}_e$,  $\ket{4}_e$ are insensitive to magnetic field fluctuations at zero magnetic field, since the average electron $\bra{i} \mathbf{S} \ket{i}$ and nuclear $\bra{i} \mathbf{I} \ket{i}$ spin moments are zero for their wavefunctions: 
\begin{align}
\bra{i} \mu_\text{B} \mathbf{B} \cdot \mathbf{g} \cdot \mathbf{S} - \mu_\text{n}  \mathbf{B}\cdot \mathbf{g}_\text{n} \cdot \mathbf{I} \ket{i} = 0.
\end{align}

For the degenerate doublet states $\ket{2,3}_g$ and  $\ket{1,2}_e$,
\begin{align}
\bra{i} \mu_\text{B} \mathbf{B} \cdot \mathbf{g} \cdot \mathbf{S} - \mu_\text{n}  \mathbf{B}\cdot \mathbf{g}_\text{n} \cdot \mathbf{I} \ket{i} &= \\
= \bra{i}  B_{\parallel} (\mu_\text{B} g_{\parallel}  S_z - \mu_\text{n}  g_\text{n} I_z )\ket{i} &= \\
 =  B_{\parallel}(\mu_\text{B}g_{\parallel} - \mu_\text{n} g_\text{n})/2,
\end{align}
leading to a linear sensitivity term for the magnetic field fluctuations along the $c$ axis $B_{\parallel}$. 

This makes the $\ket{1}_g - \ket{4}_g$ and $\ket{3}_e - \ket{4}_e$ spin transitions as well as the optical transitions connecting  $\ket{1}_g$, $\ket{4}_g$ and $\ket{3}_e$, $\ket{4}_e$ protected from the magnetic noise environment to first-order, a situation known as clock transitions.

We note that even if the spin levels $\ket{i}$, $\ket{j}$ are not sensitive to the change of the external magnetic field, the magnetic dipole of the $\ket{i}-\ket{j}$ transition is strong and is given by the dipole moment $\mu_{ij} = -\bra{i} \mu_\text{B} \mathbf{B}_\text{ac} \cdot \mathbf{g} \cdot \mathbf{S} \ket{j}/\abs{\mathbf{B}_\text{ac}}$. 
The dipole moment strongly depends on the $\mathbf{B}_\text{ac}$ magnetic field orientation.
At zero magnetic field for the $\ket{1}_g - \ket{4}_g$ spin transition, the magnetic dipole of the transition is non-zero only when  $\mathbf{B}_\text{ac}$ is oriented along the $c$ axis such that
$\mu_{1g-4g} = -\bra{1}_g \mu_\text{B} g_\parallel S_z \ket{4}_g = -\mu_\text{B}g_\parallel $. This is because the only nonzrero transition dipole moment is $\bra{1}_g S_z \ket{4}_g$, where $z$ orientation corresponds to the c-axis. For the transitions $\ket{1}_g - \ket{2}_g$ and $\ket{4}_g - \ket{2}_g$, the non-zero orientation of the ac magnetic field is along the $a$ axis such that $\mu_{1g-2g} = -\bra{1}_g \mu_\text{B} g_\perp (S_x+S_y )\ket{2}_g = -\sqrt{2} \mu_\text{B} g_\perp  $. The same is true for transitions connecting to the $\ket{3}_g$ state.

\subsection{Spin-spin dynamics}
\label{SM:flip}

The cross-relaxation processes between \ybiso{} spins strongly affects the measured coherence properties both for optical and spin transitions. The rate of the flip-flop process between spin states $\ket{0}$ and $\ket{1}$ for spin $i$ surrounded by $j$ spins can be written as \cite{Portis1956_S,Bottger2006_S},
\begin{equation}\label{eq2}
R_{\mathrm{ff}}=\frac{2\pi}{\hbar}\sum_j{\abs{\bra{0_i,1_j} \mathcal{H}_{\text{dd}}  \ket{1_i,0_j} } ^2} \frac{1}{\hbar (\Gamma^{(s)}_{\text{h}} + \Gamma^{(s)}_{\text{inh}})},
\end{equation}
where $\Gamma^{(s)}_{\text{inh}}$ ($\Gamma^{(s)}_{\text{h}}$) is the inhomogeneous (homogeneous) spin linewidth and  $\mathcal{H}_{\text{dd}}$ is the magnetic dipole-dipole interaction Hamiltonian.  
The $\mathcal{H}_{\text{dd}}$ Hamiltonian for two identical spins $S_i$ and $S_j$ is given by
\begin{align}
\mathcal{H} = \mathcal{H}_i  + \mathcal{H}_i + \mathcal{H}_{\text{dd}} = &\mathbf{I}_i \cdot \mathbf{A} \cdot \mathbf{S}_i + \mu_\text{B} \mathbf{B} \cdot \mathbf{g} \cdot \mathbf{S}_i - \mu_\text{n}  \mathbf{B}\cdot \mathbf{g}_\text{n} \cdot \mathbf{I}_i + \\ 
&+ \mathbf{I}_j \cdot \mathbf{A} \cdot \mathbf{S}_j + \mu_\text{B} \mathbf{B} \cdot \mathbf{g} \cdot \mathbf{S}_j - \mu_\text{n}  \mathbf{B}\cdot \mathbf{g}_\text{n} \cdot \mathbf{I}_j + \mathcal{H}_{\text{dd}},
\end{align}
where \cite{Abragam1970_S} (Chapter 9)
\begin{align}
\mathcal{H}_{\text{dd}} = & \frac{\mu_0}{4\pi}\mu_\text{B}^2 \abs{\Vec{r}_{ij}}^{-3}[ 
(1-3l^2) g_\perp^2 S_{ix} S_{jx} + (1-3m^2) g_\perp^2 S_{iy} S_{jy} + (1-3n^2) g_\parallel^2 S_{iz} S_{jz} \\
& - 3 l m g_\perp^2 ( S_{ix} S_{jy} + S_{iy} S_{jx}) - 3 l n g_\perp g_\parallel ( S_{ix} S_{jz} + S_{iz} S_{jx}) 
- 3 m n g_\perp g_\parallel ( S_{iy} S_{jz} + S_{iz} S_{jy})
],
\end{align}
where $\left[l, m, n\right]$ are the direction cosines of the vector $\Vec{r}_{ij}$ connecting two spins $i$ and $j$. We consider the dipole-dipole interaction $\mathcal{H}_{\text{dd}}$ as a perturbation to the Zeeman interaction allowing us to use the Fermi golden rule to calculate the cross relaxation rate using Eq. \eqref{eq2}. After integrating through finite number of crystallographic layers one can find that the flip-flop rate \cite{Car2019_S}
\begin{equation}
R_\text{ff}  = \frac{2\pi}{\hbar} \left( \frac{\mu_0}{4\pi} \mu_B^2 \right)^2 8.4 \times n_s^2 \beta_{\text{ff}} \frac{1}{\hbar (\Gamma^{(s)}_{\text{h}} + \Gamma^{(s)}_{\text{inh}})},
\end{equation}
where $n_s$ is the spatial density of $j$ spins, and $\beta_{\text{ff}}$ is the spin-spin coupling coefficient accounting for magnetic dipole-dipole interaction integrated over all directions
\begin{align}
\beta_{\text{ff}} = &\frac{1}{4\pi} \int [ (1-3l^2) g_\perp^2 S_{ix} S_{jx} + (1-3m^2) g_\perp^2 S_{iy} S_{jy} +  (1-3n^2) g_\parallel^2 S_{iz} S_{jz}- \\ &- 3 l m g_\perp^2 ( S_{ix} S_{jy} + S_{iy} S_{jx}) - 3 l n g_\perp g_\parallel ( S_{ix} S_{jz} + S_{iz} S_{jx}) - 3 m n g_\perp g_\parallel ( S_{iy} S_{jz} + S_{iz} S_{jy})] dS
\end{align}

In the case of an anisotropic magnetic interaction,  $\beta_{\mathrm{ff}}$  strongly depends on the external magnetic field orientation \cite{Zambrini2017_S}, defined by the anisotropy of the $g$ tensor. We consider the case of zero magnetic field,  where $\beta_{\mathrm{ff}}$ depends on  electron-nuclear spin wavefunctions due to the hyperfine interaction.

Let's consider the flip-flop process for the $\ket{1}_g - \ket{4}_g$ transition. In this case, since only $\bra{1}_g S_z \ket{4}_g$ is non zero, the expression for the flip-flop rate is simplified, such that
\begin{equation}
\beta_\text{ff} = \frac{1}{4\pi}\int\frac{g_\parallel^4 (l^2+m^2-2n^2)^2}{32} dS = \frac{1}{4\pi} \frac{\pi  g_\parallel^4 }{10}.
\end{equation} 

For the flip-flop process for either the $\ket{1}_g - \ket{2}_g$ or $\ket{4}_g - \ket{2}_g$ transition, or transitions involving $\ket{3}_g$, the $S_z$ component is zero, leading to a flip-flop process depending only on $g_\perp$
\begin{equation}
\beta_\text{ff} = \frac{1}{4\pi} \int\frac{g_\perp^4 (l^2+m^2-2n^2)^2}{128}dS =\frac{1}{4\pi}\frac{\pi  g_\perp^4}{40},
\end{equation} 

We can  see that the flip-flop rate is strongly dependent on electron-nuclear wavefunctions of the transition, making the flip-flop rate highly unequal. For example, for this \CWO{} crystal, the flip-flop rate for the $\ket{1}_g - \ket{4}_g$ is $0.5g_{\perp}^4/g_{\parallel}^4 \approx 100$ times slower than  for the $\ket{1}_g - \ket{2}_g$ and $\ket{1}_g - \ket{3}_g$ transitions.

For the given crystal structure the average distance between dopants can be estimated using
\begin{equation}
    r_\text{avg} = \left(\frac{V}{Zn} \right)^{1/3},
\end{equation}
where $V$ is the volume of the crystalline unit cell, $Z$ is the number of sites in the volume, and $n$ is the occupation percentage of these sites. 
The tetragonal unit cell of \CWO{} with dimension $a = b =  0.5243$ nm, $c = 1.1371$ nm gives a unit cell volume of $V = 0.2795$ nm$^3$ with an ion density of $1.43 \times 10^{22}$ cm$^{-3}$. 

The unit cell of \CWO{} contains 4 Ca$^{2+}$ ions, so that the average Ca$^{2+}$ ion-ion distance is  approximately $(V/Z)^{1/3} = 0.411$ nm. For an \ybCWO{} crystal with $\approx 5$ ppm at. doping concentration, only 0.0005\% of the Ca$^{2+}$ sites are occupied by \ybiso{}. The average distance between \ybiso{} ions then is $(V/Zn)^{1/3} = 24.2$ nm.

\subsection{Optical and spin decoherence }
\label{SM:decoh}
The observed coherence times for various optical pumping conditions allows us to estimate various contributions to optical and spin decoherence.

The optical coherence was measured using the $\ket{4}_g$ - $\ket{4}_e$  transition, which has zero sensitivity to magnetic field fluctuations in the first order.
All other allowed optical transitions contain a degenerate doublet in the ground or excited state, making them more sensitive to magnetic field noise.

The optical coherence time was measured to be $T_2^{(o)} = 0.540$ ms after reshuffling the population to avoid any optical pumping, especially at low temperatures. The optical $T_2^{(o)}$ stays constant up to 3 K but starts to drop slowly as the temperature is increased, e.g. at 4 K, $T_2^{(o)} = 0.2$ ms. The $T_2^{(o)}$ is strongly enhanced up to $T_2^{(o)}=0.75$ ms after pumping all \ybiso{} ions into the $\ket{4}_g$ state.

The effect of the optical pumping on the optical coherence suggests that the spin dynamics of \ybiso{} ions are limiting $T_2^{(o)}$ at temperature below 1~K. The spin-lattice relaxation is  very low, and the effect from magnetic field noise should be small due to the clock condition, which further suggests that spin-spin dynamics are the main limiting factor. The flip-flop process can affect the $T^{(o)}_2$ time by limiting the population lifetime of the spin ground state of optically excited  \ybiso{} ions. 
The lowest limit for the optical homogeneous linewidth can then be written as
\begin{equation}
\pi\Gamma_h^{(o)} = \frac{1}{2T_1} + \frac{1}{2}\sum_i{R_{\text{ff}}^{4_g,i_g}} + \frac{1}{2}\sum_i{R_{\text{SLR}}^{4_g,i_g}},
\end{equation}
where only population decay processes are taken into account, such as the excited state lifetime $T_1$, the spin lattice relaxation rate $R_{\text{SLR}}^{4_g,i_g}$, and the flip-flop processes $R_{\text{ff}}^{4_g,i_g}$ on  the $\ket{4}_g-\ket{i}_g$ transition, where $i\neq 4$. 

Half of the contribution to the measured homogeneous linewidth without optical pumping comes from the excited state lifetime. The second half comes from the flip-flop rate dynamics with $R_{\text{ff}} \approx 2.5\times 10^3$ s$^{-1}$. We expect that the flip-flop rate should be fast for the $\ket{4}_g-\ket{2}_g$ and $\ket{4}_g-\ket{3}_g$ transitions, and much slower for $\ket{1}_g-\ket{4}_g$ transition due to the electron-nuclear spin wavefunctions  (see previous sections). Since there are two fast processes connecting $\ket{4}_g$ state, the flip-flop rate for each of them can be estimated as  $R_{\text{ff}}^{4_g,2_g}=R_{\text{ff}}^{4_g,3_g} \approx 1.25\times 10^3$ s$^{-1}$.

A spin coherence time $T_2^{(s)}$ of 150 ms was measured for $\ket{1}_g-\ket{4}_g$ transition after pumping all the \ybiso{} ions into $\ket{1}_g$ level. The limit for the spin homogeneous linewidth is 
\begin{equation}
\pi\Gamma_h^{(s)} = \frac{1}{2}\sum_{i,j}{R_{\text{ff}}^{j_g,i_g}} + \frac{1}{2}\sum_{i,j}{R_{\text{SLR}}^{j_g,i_g}},
\end{equation}
when SLR and flip-flop processes can happen from both $\ket{1}_g$, $\ket{4}_g$ states such that in the summation  $j={1,4}$ and $i\neq j$.
At 100~mK temperatures, the SLR rate is much slower than the measured decoherence (see Section 4.E), making its contribution negligible. When polarising the whole spin ensemble in the $\ket{1}_g$ state, some flip-flop processes are quenched, such that the $\Gamma_h^{(s)}$ is limited by
\begin{equation}
\pi\Gamma_h^{(s)} = \frac{1}{2}({R_{\text{ff}}^{1_g,4_g} + R_{\text{ff}}^{4_g,1_g}}),
\end{equation}
corresponding to the flip-flop processes involving spins excited via the spin echo sequence on the $\ket{1}_g-\ket{4}_g$ transition. In our experiments, we manipulate only a small part of the optical inhomogeneous broadening around 1~MHz, exciting around 0.5\% of the population. In this case, only the processes of flipping from $\ket{4}_g$ to $\ket{1}_g$ will contribute such that
\begin{equation}
\pi\Gamma_h^{(s)} = \frac{1}{2}R_{\text{ff}}^{1_g,4_g}.
\end{equation}
This allows us to give the lower bound of $R_{\text{ff}}^{1_g,4_g} = 13.3$ s$^{-1}$ for the flip-flop process on the $\ket{4}_g - \ket{1}_g$ transition assuming the spin polarisation is perfect. 

\subsection{Electron paramagnetic resonance (EPR) measurements}
\label{SM:epr}

CW-EPR measurements were performed using a Bruker EMX spectrometer operating in the X-band with microwave frequencies around 9.4 GHz. The spectrometer is equipped with a He-flow cryostat (ESR900) and a cryogen-free cooler (Bruker Stinger) going down to 7 K. The angular dependence of EPR with respect to the static field was measured using an automatic goniometer installed on the spectrometer. Angular measurements were done in two planes, one containing the $c$ axis (denoted as the $c-a$ plane) and one perpendicular to the $c$ axis ($a-b$ plane). The 0$^{\circ}$ angle in the $c-a$ plane corresponds to the direction parallel to the $c$ axis. The microwave power was set low ($<1$~mW) to prevent the EPR signal from saturating. The field modulation was kept under 1 G to avoid  distortion of the EPR lines due to over-modulation effects.

\begin{figure*}[hbtp!]
	\includegraphics[width=0.76\linewidth, trim=0.cm 0.0cm 0.cm 0.0cm,clip]{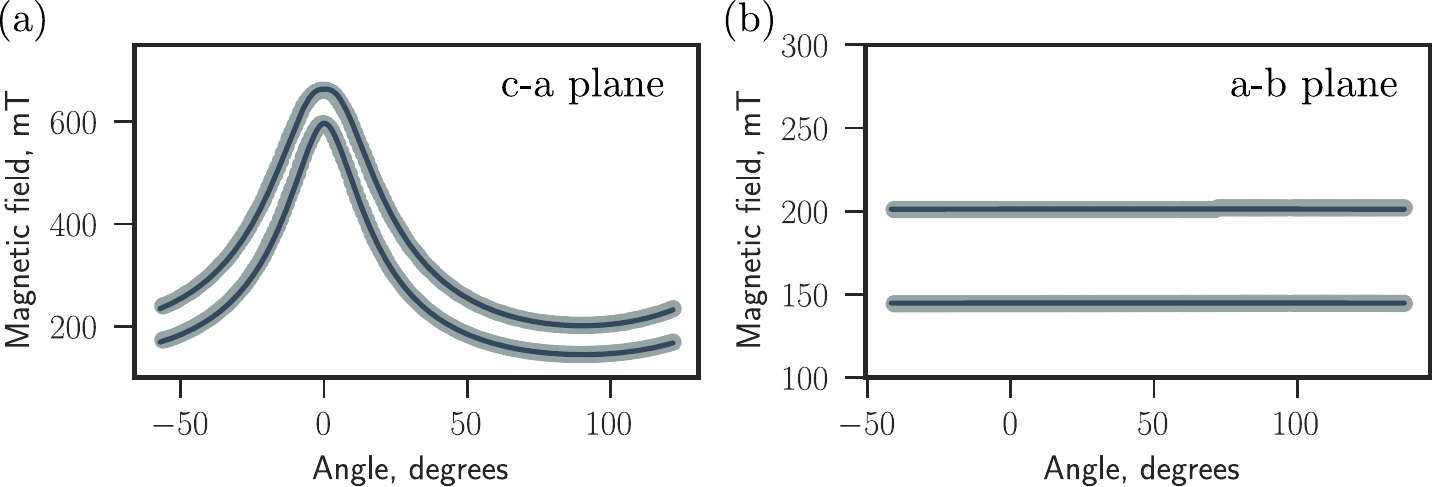}
	\caption{ 
	(color online) \textbf{EPR angular spectra}. EPR transitions for \ybCWO{} in \dfs{}(0) ground state measured as a function of the angle in the c-a and a-b planes for tetragonal site (a)-(b). Grey points: experimental data; lines: spin Hamiltonian model.
}
	\label{SMfig:epr}
\end{figure*}

\begin{figure*}[hbtp!]
	\includegraphics[width=0.6\linewidth, trim=0.cm 0.0cm 0.cm 0.0cm,clip]{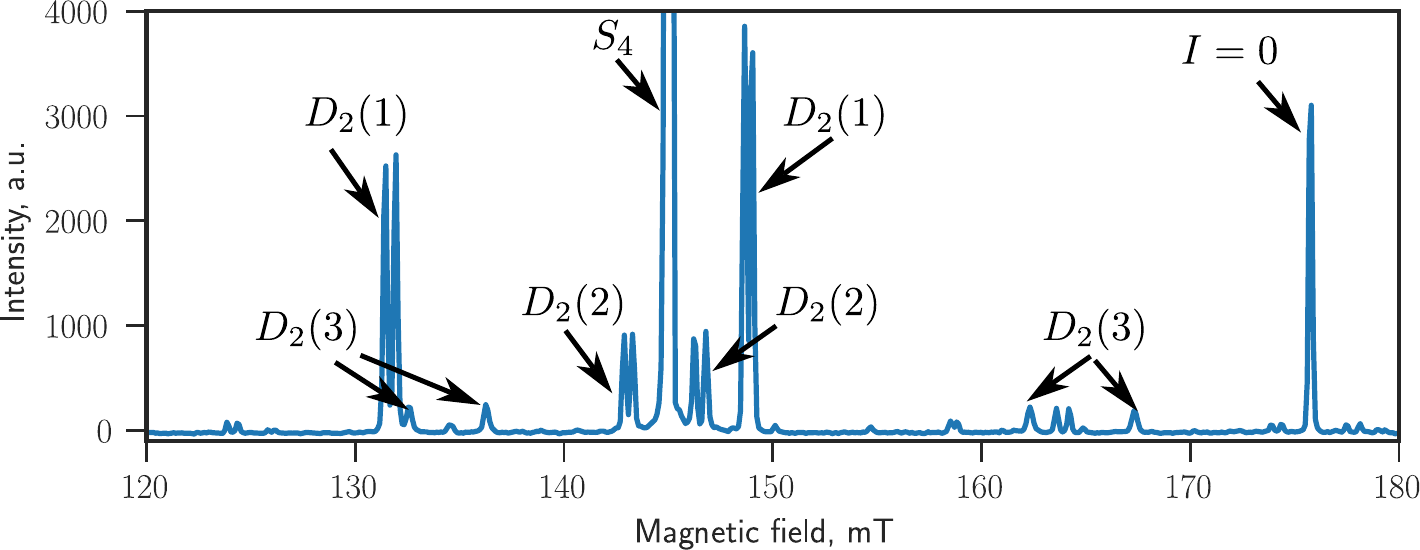}
	\caption{ 
	(color online) EPR transitions for \ybCWO{} in the \dfs{}(0) ground state measured with a magnetic field orientation with a small angle with respect  to the $a$-axis. The lines corresponding to the tetragonal $S_4$ site (peaking outside the graph) and the three orthorhombic $D_2$ sites  for \ybiso{} ions are indicated, as well as the line corresponding to $I=0$ \yb{} isotopes. The second part of the $S_4$ doublet appears at 201 mT.
}
	\label{SMfig:eprtrace}
\end{figure*}

Angular scans of different EPR lines in the $c-a$ and  $a-b$ planes are shown in \figref{SMfig:epr}. Several lines with different intensities were observed in the angular scans. We focused on extracting the lines forming doublets, which should come from the $I = 1/2$ \ybiso{}  isotope. The most intense line comes \ybiso{} ions occupying the tetragonal $S_4$ site of Ca$^{2+}$ ions and dominates the spectra of \figref{SMfig:epr}(a-b). It shows nearly no angular dependence in the $a-b$ plane orthogonal to the $c$ axis in agreement with the symmetry of the spin Hamiltonian. The resulting parameters for $g$ and $A$ extracted from previous EPR measurements \cite{Sattler1970_S} explain perfectly the observed spectra. The spin Hamiltonian parameters are summarized in \tabref{SM:eprTable}.
The other lines which correspond to the \yb{} isotopes with nuclear spin $I=0$ are visible with around 5\% of intensity with respect to the \ybiso{} isotope (\figref{SMfig:eprtrace}). 
Narrow spin resonance lines with widths of few a MHz have been measured, making it possible to couple to high-quality microwave resonators.

In addition to the tetragonal $S_4$ site, contributions from lower symmetry orthorhombic $D_2$ sites are clearly visible in the spectra but with much lower intensity (\figref{SMfig:eprtrace}). These $D_2$ sites were previously observed in EPR studies \cite{Ranon1964_S} and  attributed to nearby charge compensation centers giving rise to an added asymmetry of the crystal field configuration at the \yb{} site. 

\setlength{\tabcolsep}{0.1em} 
\renewcommand{\arraystretch}{1.25}
\begin{table*}[!htb]
 \caption{ Spin Hamiltonian parameters for the tetragonal $S_4$ site (ground \dfs{}(0) and excited \dfc{}(0) states manifolds) 
 in \ybCWO{} including the $g$ and $A$ tensor parameters.
 }
\begin{ruledtabular}
\begin{minipage}[t]{.6\linewidth}
 \begin{tabular}{cc cc cc }
  Site      & $n$, ppm & $g_\perp$ & $g_\parallel$ &  $A_\perp$, GHz & $A_\parallel$, GHz \\ 
$S_4$ (gnd \dfs{}(0)) & 4.96 &  3.916 & 1.053 &  3.08384(1) & 0.787(1)\\
$S_4$ (exc \dfc{}(0)) &  & 1.362(1)   & -1.452(1)  &  2.734(2) & -2.878(2) \\
 \end{tabular}
\end{minipage}
\label{SM:eprTable}
\end{ruledtabular}
 \end{table*}

\subsection{Concentration estimation}
\label{SM:conc}

The EPR measurements were used to estimate the concentration of sites occupied by  \ybiso{} ions in the \CWO{} crystal with a nominal concentration of 20 ppm. For this, comparison with a previously characterised \CWO{} sample was used by performing measurements under the same conditions. The reference sample was used for the single-spin detection experiments from \cite{Wang2023_S} was used. This gave 4.96 ppm concentration for the main tetragonal $S_4$ site from the nominal concentration of 20 ppm. The intrinsic \yb{} concentration in undoped \CWO{} crystals is estimated to be a few ppb, based on previous growth and characterization \cite{Billaud2025_S}, which is negligible compared to the doping level used in this work. 

Around 37\% of the nominal doping concentration is present in the crystal. This is significantly higher compared with a previous report on the distribution coefficient for \CWO{} crystal, which for Yb$^{3+}$ was estimated to be around 0.15-0.2 \cite{Nassau1963_S}. Around 70\% of the ions are located in the main tetragonal $S_4$ site. 
Charge compensation methods based on adding sodium (Na) or niobium (Nb) oxides during the crystal growth significantly increase the occupation probabilities, but at higher doping concentrations \cite{Nassau1963_S}. These effects at lower doping concentrations and their effect on optical and spin properties still require investigation.

\section{State representations and selection rules}
\label{SM:selection}

In this section, we derive the optical transition selection rules and state assignment for the \ybCWO{} crystal. The main idea  is to follow the group theory approach to explain the optical absorption spectra and optical branching ratios quantified in the previous section. We show that the symmetry of the crystal and the observed transition rules appropriately explain the observed behavior.

We  write down the selection rules for the electric and magnetic dipole operators using the rules between  irreducible representations. We then figure out the corresponding labels for the observed states in our system by measuring which optical transitions are observed in the crystal for different orientations of the light polarization relative to the crystal symmetry axes. We measured absorption with light polarisation along the $c$ axis ($\pi$), perpendicular to the $c$ axis ($\sigma$), and with  light propagating along the $c$ axis ($\alpha$).
 
Yb$^{3+}$ ions substitute Ca$^{2+}$ ions in the sites with tetragonal $S_4$ point symmetry in the \CWO{} crystal. However, previously, the crystal symmetry for rare-earth ions in scheelite crystals was described by $D_{2d}$ symmetry due to a small distortion from $D_{2d}$ to $S_4$ \cite{Trabelsi2011_S}. The crystal field Hamiltonian for tetragonal symmetry can be written in terms of the Stevens operator equivalents as
$$
H_{CF} = B_2^0 O_2^0 + B_4^0 O_2^0+ B_6^0 O_6^0 +B_4^4 O_2^4 + B_6^4 O_6^6,
$$
where $B_k^q$ (where $k = 2, 4, 6$ $|q |\leq k$) are the crystal-field parameters. 
The character table for $S_4$ and $D_{2d}$ symmetries are given in \tabref{SM:char} \cite{Koster1963_S,Powell2010_S}.

\setlength{\tabcolsep}{0.4em} 
\renewcommand{\arraystretch}{1.25}
\begin{table}[!htb]
\caption{Character table for $S_4$ and $D_{2d}$ symmetry, where $\omega=e^{i\pi/4}$.}
\begin{ruledtabular}
\begin{minipage}[t]{.4\linewidth}
\begin{tabular}{c|c|cccc}
$S_4$  &   & $E$ & $C_2$ & $S_4$  & $\bar{S_4}$  \\ \hline
$A$& $\Gamma_1$  & +1 & +1 & +1        & +1      \\
$B$ &$\Gamma_2$ &  +1 & +1 & -1        & -1     \\
 &$\Gamma_{3}$ &  +1  & -1 &  -i         & -i   \\
 &$\Gamma_{4}$ &  +1  & -1 & i         & i   \\
 & $\Gamma_{5}$ &  +1  & i & -$\omega^3$         & $\omega$ \\
 & $\Gamma_{6}$ &  +1  & -i  & $\omega$        & $-\omega^3$ \\
 & $\Gamma_{7}$ &  +1  & i & $\omega^3$         & -$\omega$ \\
 & $\Gamma_{8}$ &  +1  & -i & -$\omega$         & $\omega^3$  \\
\end{tabular}
\label{SM:char}
\end{minipage}
\end{ruledtabular}

\vspace{10pt}
\begin{ruledtabular}
\begin{minipage}[t]{.6\linewidth}
\begin{tabular}{c|c|cccccccccc}
$D_{2 d}$ & & $E$ & $2 S_4$ & $C_2$ & $2 C_2^{\prime}$ & $2 \sigma_d$ & $R$ & $2 R S_4$ & $R C_2$ & $2 R C_2^{\prime}$ & $2 R \sigma_d$ \\
\hline$A_1$ & $\Gamma_1$ & 1 & 1 & 1 & 1 & 1 & 1 & 1 & 1 & 1 & 1 \\
$A_2$& $\Gamma_2$ & 1 & 1 & 1 & -1 & -1 & 1 & 1 & 1 & -1 & -1 \\
$B_1$& $\Gamma_3$ & 1 & -1 & 1 & 1 & -1 & 1 & -1 & 1 & 1 & -1 \\
$B_2$& $\Gamma_4$ & 1 & -1 & 1 & -1 & 1 & 1 & -1 & 1 & -1 & 1 \\
$E$& $\Gamma_5$ & 2 & 0 & -2 & 0 & 0 & 2 & 0 & -2 & 0 & 0 \\
$D_{1 / 2}$& $\Gamma_6$ & 2 & $\sqrt{2}$ & 0 & 0 & 0 & -2 & $-\sqrt{2}$ & 0 & 0 & 0 \\
${ }_2 S$& $\Gamma_7$ & 2 & $-\sqrt{2}$ & 0 & 0 & 0 & -2 & $\sqrt{2}$ & 0 & 0 & 0 \\

\end{tabular}
\end{minipage}
\end{ruledtabular}

\end{table}

The 4f$^{13}$ configuration of Yb$^{3+}$ consists of only two electronic multiplets: \dfc{} in the ground state and \dfs{} in the excited state. We then  write the spin-orbit \dfc{} and \dfs{} multiplets in terms of irreducible representations of $S_4$ and $D_{2d}$ symmetry. We use the corresponding full-rotational group compatibility table (\tabref{SM:irrep}) for a free-ion for both the $J=7/2$ and $J=5/2$ levels \cite{GrollerWalrand1996_S}.

\setlength{\tabcolsep}{0.4em} 
\renewcommand{\arraystretch}{1.25}
\begin{table}[!htb]

\caption{Irreducible representations for $S_4$ and $D_{2d}$ symmetry.}
\begin{ruledtabular}
\begin{minipage}[t]{.3\linewidth}
\begin{tabular}{l|llll|lll}
$J$ & $S_4$ \\  \hline
$1/2$   & $ \Gamma_{7,8}$ \\
$5/2$   & $ 2\Gamma_{5,6} + \Gamma_{7,8}$ \\
$7/2$ & $ 2\Gamma_{5,6} + 2\Gamma_{7,8}$    
\end{tabular}
\end{minipage}
\end{ruledtabular}
\begin{ruledtabular}
\begin{minipage}[t]{.3\linewidth}
\begin{tabular}{l|llll|lll}
$J$ &  $D_{2d}$ \\ \hline
$1/2$   & $ \Gamma_{7}$ \\
$5/2$   & $ 2\Gamma_{6} + \Gamma_{7}$ \\
$7/2$ & $ 2\Gamma_{6} + 2\Gamma_{7}$    
\end{tabular}
\label{SM:irrep}
\end{minipage}
\end{ruledtabular}
\end{table}

The \dfs{} and \dfc{} multiplets split into 4 and 3 doublets, respectively, corresponding to $\Gamma_{5,6}$ or $\Gamma_{7,8}$  irreps for $S_4$ or $\Gamma_{6}$ or $\Gamma_{7}$ for $D_{2d}$. 

\subsection{Ground state doublet \dfs{}(0)}
Previously, the first doublet of the $^2$F$_{7/2}$ multiplet was attributed to $\Gamma_{5,6}$ irrep for $S_4$ symmetry based on  EPR measurements \cite{Ranon1964_S}. Due to the $S_4$ ($D_{2d}$) site symmetry, nonzero crystal field parameters correspond to $B_2^0$, $B_4^0$, $B_6^0$, $B_4^4$ and $B_6^4$, strongly limiting the mixing between different $\ket{J,M_J}$ states. The wavefunction components $\ket{J, M_J}$ are mixed by  $B_4^4$ and $B_6^4$ terms and are given by \cite{Sattler1970_S} 
$$ \ket{\Gamma_5} = a\ket{7/2,+5/2} + b\ket{7/2,-3/2}, 
\ket{\Gamma_6} = a^*\ket{7/2,-5/2} + b^*\ket{7/2,+3/2},
$$
$$ \ket{\Gamma_7} = c\ket{7/2,-7/2} + d\ket{7/2,+1/2},
\ket{\Gamma_8} = -c^*\ket{7/2,7/2} - d^*\ket{7/2,-1/2}.
$$

Based on these wavefunctions, one can calculate the g-factors of the various doublets represented by $\Gamma_{\pm}$ irreps as 
$$ g_{\parallel} = 2g_J\mel**{\Gamma_+}{J_z}{\Gamma_+},
 g_{\perp} = g_J(\mel**{\Gamma_+}{J_x}{\Gamma_-}+\mel**{\Gamma_-}{J_x}{\Gamma_+}),
$$
with $g_J = 1 + (J(J+1) + S(S+1) - L(L+1))/(2J(J+1))$, such that $g_{\frac{5}{2}} = 6/7$ and $g_{\frac{7}{2}} = 8/7$ with $L=3$ and $S=1/2$, and 
$$ \bra{J,M'_J} J_z \ket{J,M_J} = \delta_{M'_J,M_J} M_J,
$$
$$ 
\bra{J,M'_J} J_x \ket{J,M_J} = (\delta_{M'_J,M_J+1}+\delta_{M_J,M'_J+1})\frac{1}{2}(J(J+1) - M_JM'_J)^{1/2}.
$$
Following the results from \cite{Sattler1970_S}, we get the expressions for $g_{\parallel}$ and $g_{\perp}$
for \dfs{} 
$$
\ket{\Gamma_{5,6}}: \text{ } g_{\parallel} = g_{7/2}(5\abs{a}^2 - 3\abs{b}^2), \text{ } g_{\perp} = 2\sqrt{3}g_{7/2}(ab+a^*b^*),
$$
$$
\ket{\Gamma_{7,8}}: \text{ } g_{\parallel} = g_{7/2}(7\abs{c}^2 - \abs{d}^2), \text{ } g_{\perp} = - 2g_{7/2}(d^2+d^{*2}),
$$
resulting in the following relations when $\Gamma_{5}$ ($\Gamma_{7}$) is above $\Gamma_{6}$ ($\Gamma_{8}$) in energy
$$
\ket{\Gamma_{5,6}}\text{ : } 4g_{\perp}^2 = -3g_{\parallel}^2 + 6g_{\frac{7}{2}}g_{\parallel}+45g_{\frac{7}{2}}^2,
$$
$$
\ket{\Gamma_{7,8}}\text{ : } 2g_{\perp} = -g_{\parallel} - 7g_{\frac{7}{2}}.
$$

When $\Gamma_{5}$ ($\Gamma_{7}$) is below $\Gamma_{6}$ ($\Gamma_{8}$) in energy giving $g_\parallel<0$
$$
\ket{\Gamma_{5,6}}: \text{ } g_{\parallel} = g_{7/2}(-5\abs{a}^2 + 3\abs{b}^2), \text{ } g_{\perp} = 2\sqrt{3}g_{\frac{7}{2}}(ab+a^*b^*),
$$
$$
\ket{\Gamma_{7,8}}: \text{ } g_{\parallel} = g_{7/2}(-7\abs{c}^2 + \abs{d}^2), \text{ } g_{\perp} = - 2g_{7/2}(d^2+d^{*2}),
$$
and  
$$
\ket{\Gamma_{5,6}}\text{ : } 4g_{\perp}^2 = -3g_{\parallel}^2 - 6g_{\frac{7}{2}}g_{\parallel}+45g_{\frac{7}{2}}^2,
$$
$$
\ket{\Gamma_{7,8}}\text{ : } 2g_{\perp} = g_{\parallel} - 7g_{\frac{7}{2}}.
$$

In $D_{2d}$ symmetry, $\Gamma_6$ and $\Gamma_7$ correspond to $\Gamma_{5,6}$ and $\Gamma_{7,8}$, respectively. These expressions are written for $S_4$ symmetry but are also valid for $D_{2d}$ irreps. As derived previously \cite{Ranon1964_S}, the first expression for $\ket{\Gamma_{5,6}}$ explains well the experimental values $g_{\parallel} = 1.05$ and $g_{\parallel} = 3.92$ for the ground state \dfs{}, with $a=0.700$ and $b=0.714$. So, we attribute the ground state doublet \dfs{}(0) to the $\ket{\Gamma_{5,6}}$ irreps in $S_4$ ($\ket{\Gamma_{6}}$ in $D_{2d}$).

\subsection{Excited state doublet \dfc{}(0)}
For \dfc{}, we get similar expressions. Since two $\Gamma_{5,6}$ and one $\Gamma_{7,8}$  are expected for $J=5/2$ we can write 
$$ \ket{\Gamma_5} = a\ket{5/2,+5/2} + b\ket{5/2,-3/2}, 
\ket{\Gamma_6} = -a^*\ket{5/2,-5/2} - b^*\ket{5/2,+3/2},
$$
$$ \ket{\Gamma_7} = \ket{5/2,-1/2},  
\ket{\Gamma_8} = \ket{5/2,+1/2},
$$
$$
\ket{\Gamma_{5,6}}: \text{ } g_{\parallel} = g_{\frac{5}{2}}(5\abs{a}^2 - 3\abs{b}^2), \text{ } g_{\perp} = -\sqrt{5}g_{\frac{5}{2}}(ab^*+ba^*),
$$
$$
\ket{\Gamma_{7,8}}: \text{ } g_{\parallel} = g_{5/2}, \text{ } g_{\perp} = 3g_{5/2},
$$
resulting in following relations for $\Gamma_{5}$ ($\Gamma_{7}$) above $\Gamma_{6}$ ($\Gamma_{8}$) in energy
$$
\ket{\Gamma_{5,6}}\text{ : } 16g_{\perp}^2 = -5g_{\parallel}^2 + 10g_{\frac{5}{2}}g_{\parallel}+75g_{\frac{5}{2}}^2,
$$
$$
\ket{\Gamma_{7,8}}\text{ : } g_{\perp} = -3g_{\parallel}.
$$

When $\Gamma_{5}$ ($\Gamma_{7}$) is below $\Gamma_{6}$ ($\Gamma_{8}$) in energy, $g_\parallel<0$, and
$$\ket{\Gamma_{5,6}}: \text{ } g_{\parallel} = g_{\frac{5}{2}}(-5\abs{a}^2 + 3\abs{b}^2), \text{ } g_{\perp} = -\sqrt{5}g_{\frac{5}{2}}(ab^*+ba^*),
$$
$$
\ket{\Gamma_{7,8}}: \text{ } g_{\parallel} = g_{5/2}, \text{ } g_{\perp} = 3g_{5/2},
$$
resulting in following relations
$$
\ket{\Gamma_{5,6}}\text{ : } 16g_{\perp}^2 = -5g_{\parallel}^2 - 10g_{\frac{5}{2}}g_{\parallel}+75g_{\frac{5}{2}}^2,
$$
$$
\ket{\Gamma_{7,8}}\text{ : } g_{\perp} = 3g_{\parallel}.
$$

Again, in $D_{2d}$ symmetry, the situation is equivalent with $\Gamma_6$ and $\Gamma_7$ corresponding to $\Gamma_{5,6}$ and $\Gamma_{7,8}$, respectively. Previously, the values for the lowest Kramers doublet in the \dfc{} excited state were measured to be $g_\perp = 1.5$ and $g_\parallel = -1.2$ \cite{Jones1967_S}. For $B \perp c$, this gives a g-factor $g_{\text{eff}} = 2.4$ and 5.2, and for $B \parallel c$ $g_{\text{eff}} = 0.15$ and 2.23. They also concluded that the lowest doublet has a $\Gamma_{5,6}$ irrep based on polarisation properties of the different crystal field levels absorption spectra \cite{Jones1967_S}. A negative $g_\parallel$ value was assumed such that the expression for g-factors is then
$$
\ket{\Gamma_{5,6}}\text{ : } 16g_{\perp}^2 = -5g_{\parallel}^2 - 10g_{\frac{5}{2}}g_{\parallel}+75g_{\frac{5}{2}}^2,
$$
such that for $g_\parallel = -1.44$ observed in this work, it predicts $g_\perp^{th} = 1.42$, while $g_\perp = 1.29$ was measured. The deviation can come from the $J=7/2$ mixing to the excited state wavefunction.

\subsection{Effect of $J$ mixing}
The effect of $J$ mixing with $J=7/2$ in the can be derived assuming wavefunctions
$$ \ket{\Gamma_5} = a\ket{5/2,+5/2} + b\ket{5/2,-3/2} +c\ket{7/2,+5/2} + d\ket{7/2,-3/2}, 
$$
$$
\ket{\Gamma_6} = -a^*\ket{5/2,-5/2} - b^*\ket{5/2,+3/2}+c^*\ket{5/2,-5/2} + d^*\ket{5/2,+3/2},
$$
Following the derivation from \cite{Sattler1970_S}, the  expressions of $g_{\parallel}$ and $g_\perp$ are
$$
\text{ } g_{\parallel} = g_{\frac{5}{2}}(5\abs{a}^2 - 3\abs{b}^2)-\frac{2\sqrt6}{7}(a^*c + ac^*)
-\frac{2\sqrt{10}}{7}(b^*d + bd^*)
+g_{\frac{7}{2}}(5\abs{c}^2 - 3\abs{d}^2), 
$$
$$
\text{ } g_{\perp} = \abs{-2\sqrt{5}g_{\frac{5}{2}}ab -\frac{2\sqrt{30}}{7}bc -\frac{2\sqrt{2}}{7}ad + 4\sqrt{3}g_{\frac{7}{2}}cd}.
$$

The observed $g_\parallel$ and $g_\perp$ values for the excited state can be explained with the mixing of $R = 4$\%, where $R=(\abs{c}^2+\abs{d}^2)/(\abs{a}^2+\abs{b}^2)$ is the ratio between two $J$ components of the wavefunction.

\subsection{Selection rules}

We check the set of irreps for each multiplet \dfs{}(0) and \dfc{}(0) using the observed polarisation properties. 
From the measured optical absorption, we got a maximum absorption for the $\alpha$ ($k \parallel c$) and $\sigma$ ($E \perp c$) polarisations, while the $\pi$ polarisation gave a 7 times lower absorption. We then calculate the selection rules between different states. For this, we use the multiplication table for $S_4$ and $D_{2d}$ symmetries  (\tabref{SM:multS4} and \tabref{SM:multD2d}) to get the transition rules between $\Gamma_i$ and $\Gamma_j$, and we check if electric or magnetic dipoles (ED and MD) connect them using the corresponding table (\tabref{SM:EDMD}) \cite{GrollerWalrand1996_S}.

The calculated selection rules between the crystal field levels are shown in \tabref{SM:EDMD}. We see that electric and magnetic dipoles can explain the different line polarisations, making it difficult to decide on the dipole nature and how to connect irreps. Moreover, both electric and magnetic dipoles are expected to appear in the measured spectra since $\Delta J = \pm 1$, suggesting that additional analysis involving the nuclear spin states is needed.

We add that the nuclear spin 1/2 has the irrep $\Gamma_{7,8}$ according to \tabref{SM:irrep}. Using the multiplication table, the resulting irreps with the nuclear spin for $S_4$ are given by
\begin{align}
    \Gamma_{5,6} \times \Gamma_{7,8} = 2\Gamma_2 + \Gamma_3 + \Gamma_4, \\
    \Gamma_{7,8} \times \Gamma_{7,8} = 2\Gamma_1 + \Gamma_3 + \Gamma_4,
\end{align}
while for $D_{2d}$ 
\begin{align}
    \Gamma_{6} \times \Gamma_{7} = \Gamma_3 + \Gamma_4 + \Gamma_5, \\
    \Gamma_{7} \times \Gamma_{7} = \Gamma_1 + \Gamma_2 + \Gamma_5.
\end{align}
Introducing the nuclear spin into the ground and excited states splits them into three levels, where $\Gamma_{3,4}$ for $S_4$ ($\Gamma_5$ for $D_{2d}$) correspond to doubly degenerate energy levels in each manifold. This allows us to get the transition selection rules between all the states. Using electric and magnetic transition dipole moments (ED and MD) shown in \tabref{SM:EDMD}, we get transition rules for each choice of irreps between the ground and excited manifolds for $S_4$ (\tabref{SM:BRtabS4}) and $D_{2d}$ (\tabref{SM:BRtabD2d}).

Comparing the observed (\tabref{SM:BRtab}) and predicted transition rules for $S_4$ (\tabref{SM:BRtabS4}) and $D_{2d}$ (\tabref{SM:BRtabD2d}), we make several observations:
\begin{itemize}
    \item The measured transition rules for $\alpha$ polarisation are well explained for both symmetries, although they do not show any difference difference depending on the choice of the dipole and the irrep.
    \item The $\pi$ polarisation measurements, for which absorption is 7 times lower, can only be explained by adding  MD transitions. This is the case for both symmetries. When assuming that the irrep for ground and excited states are different (left \tabref{SM:BRtabS4} and \tabref{SM:BRtabD2d}), the spectra are expected to have more ED transitions, which are not experimentally observed. When assuming the irrep is the same between the ground and excited state (right \tabref{SM:BRtabS4} and \tabref{SM:BRtabD2d}), only one ED-allowed $\pi$ transition does not appear in the observed spectra $\ket{2,3}_g  - \ket{1,2}_e$. While not excluding the opposite, this suggests that the irrep stays the same for ground and excited states. This is true for both symmetries.
    \item For $\sigma$ polarisation, the situation is different. The absorption for the D and E lines with $\sigma$ polarisation can be explained only using MD with the right Tables when assuming that the irrep does not change. This is true for both symmetries. For $S_4$ symmetry, the absorption for MD-allowed $\ket{1}_g - \ket{4}_e$ and $\ket{4}_g - \ket{3}_e$ $\sigma$ transitions are theoretically allowed, but not experimentally observed. However, for $D_{2d}$ symmetry, it is possible to find an attribution of irreps, where these transitions are forbidden, and D and E lines are still well explained. This suggests that $D_{2d}$ symmetry  fits better to explain the spectra.
\end{itemize}

The predicted transition rules, in general, explain the observed absorption spectra satisfactorily. When assuming $D_{2d}$ symmetry and that the irrep does not change between two manifolds (right table in \tabref{SM:BRtabD2d}), the predicted transition rules can explain the observed absorption spectra, with only a single exception of an ED-allowed $\pi$ transition not appearing in the spectra. The physical origin of this is not yet fully understood and will require further investigation, including identification of the irreducible representations of all CF levels or the use of other rare-earth dopants.  

EPR measurements of $g$ factors do not contradict  the choice of $D_{2d}$ symmetry, suggesting that the ground and excited state manifolds have $\Gamma_6$ irreps.
The extracted hyperfine $A$-tensor has $A_\perp = -2.72$ GHz and $A_\parallel = -2.87$ GHz with the ratio between the two elements close to 1.

\setlength{\tabcolsep}{0.4em} 
\renewcommand{\arraystretch}{1.25}
\begin{table*}[!htb]
\caption{Selection rules in $S_4$ (left) and $D_{2d}$ (right) symmetry for electric and magnetic dipoles (\red{red}).}
\begin{ruledtabular}
\begin{minipage}[t]{.35\linewidth}
    \begin{tabular}{c|cc}
$S_4$       & $\Gamma_{5,6}$            & $\Gamma_{7,8}$            \\ \hline
$\Gamma_{5,6}$ & $\alpha$, $\sigma$  (\red{$\alpha$, $\sigma$, $\pi$})      & $\alpha$, $\sigma$, $\pi$ (\red{$\alpha$, $\pi$})  \\
$\Gamma_{7,8}$ & $\alpha$, $\sigma$, $\pi$ (\red{$\alpha$, $\pi$})& $\alpha$, $\sigma$   (\red{$\alpha$, $\sigma$, $\pi$})           
\end{tabular}
\end{minipage}%
\end{ruledtabular}
\begin{ruledtabular}
\begin{minipage}[t]{.35\linewidth}
    \begin{tabular}{c|cc}
 $D_{2d}$      & $\Gamma_{6}$            & $\Gamma_{7}$            \\ \hline
$\Gamma_{6}$ & $\alpha$, $\sigma$  (\red{$\alpha$, $\sigma$, $\pi$})      & $\alpha$, $\sigma$, $\pi$ (\red{$\alpha$, $\pi$})  \\
$\Gamma_{7}$ & $\alpha$, $\sigma$, $\pi$ (\red{$\alpha$, $\pi$})& $\alpha$, $\sigma$   (\red{$\alpha$, $\sigma$, $\pi$})           
\end{tabular}
\end{minipage}%
\end{ruledtabular}

\vspace{10pt}
\begin{ruledtabular}
\begin{minipage}[t]{.35\linewidth}
    \begin{tabular}{c|ccc}
  $S_4$      & $\Gamma_{1}$    & $\Gamma_{2}$     & $\Gamma_{3,4}$   \\ \hline
$\Gamma_{1}$  & - (\red{$\sigma$}) & $\pi$  (\red{-})  & $\alpha$, $\sigma$ (\red{$\alpha$, $\pi$})  \\
$\Gamma_{2}$  & $\pi$ (\red{-}) & - (\red{ $\sigma$})       & $\alpha$, $\sigma$ (\red{$\alpha$, $\pi$}) \\
$\Gamma_{3,4}$ & $\alpha$, $\sigma$ (\red{$\alpha$, $\pi$}) & $\alpha$, $\sigma$ (\red{ $\alpha$, $\pi$}) & $\pi$  (\red{$\sigma$})            
\end{tabular}
\end{minipage}%
\end{ruledtabular}
\begin{ruledtabular}
\begin{minipage}[t]{.57\linewidth}
    \begin{tabular}{c|ccccc}
 $D_{2d}$       & $\Gamma_{1}$    & $\Gamma_{2}$  &  $\Gamma_{3}$    & $\Gamma_{4}$     & $\Gamma_{5}$   \\ \hline
$\Gamma_{1}$  & - (\red{-}) &  - (\red{$\sigma$})  &  - (\red{-})  &  $\pi$ (\red{-}) &  $\alpha$, $\sigma$ (\red{$\alpha$, $\pi$})  \\
$\Gamma_{2}$  & - (\red{$\sigma$}) &  - (\red{-})  &  $\pi$ (\red{-})  &  - (\red{-}) &  $\alpha$, $\sigma$ (\red{$\alpha$, $\pi$})  \\
$\Gamma_{3}$ & - (\red{-}) &  $\pi$ (\red{-})  &  - (\red{-})  &  - (\red{$\sigma$}) &  $\alpha$, $\sigma$ (\red{$\alpha$, $\pi$})  \\
$\Gamma_{4}$  & $\pi$ (\red{-}) &  - (\red{-})  &  - (\red{$\sigma$})  &  - (\red{-}) & $\alpha$, $\sigma$ (\red{$\alpha$, $\pi$}) \\
$\Gamma_{5}$ & $\alpha$, $\sigma$ (\red{$\alpha$, $\pi$}) &  $\alpha$, $\sigma$ (\red{$\alpha$, $\pi$})  &  $\alpha$, $\sigma$ (\red{$\alpha$, $\pi$})  &  $\alpha$, $\sigma$ (\red{$\alpha$, $\pi$}) &  $\pi$ (\red{$\sigma$})  \\
\end{tabular}
\end{minipage}%
\end{ruledtabular}
\label{SM:EDMD}
\end{table*}

\setlength{\tabcolsep}{0.4em} 
\renewcommand{\arraystretch}{1.25}
\begin{table}[!htb]
\caption{Multiplication table for $S_4$ symmetry \cite{Koster1963_S}.}
\begin{ruledtabular}
\begin{minipage}[t]{.4\linewidth}
\begin{tabular}{cccccccc|c}
$\Gamma_1$ & $\Gamma_2$ & $\Gamma_3$ & $\Gamma_4$  & $\Gamma_5$ & $\Gamma_6$ & $\Gamma_7$ & $\Gamma_8$  &   \\ \hline
$\Gamma_1$ & $\Gamma_2$ & $\Gamma_3$  & $\Gamma_4$ & $\Gamma_5$ & $\Gamma_6$ & $\Gamma_7$  & $\Gamma_8$ & $\Gamma_1$ \\
  & $\Gamma_1$ & $\Gamma_4$ & $\Gamma_3$  & $\Gamma_7$  & $\Gamma_8$ & $\Gamma_5$ & $\Gamma_6$ & $\Gamma_2$ \\
  &   & $\Gamma_2$ & $\Gamma_1$ & $\Gamma_8$ & $\Gamma_5$ & $\Gamma_6$ & $\Gamma_7$  & $\Gamma_3$  \\
  &   &   & $\Gamma_2$ & $\Gamma_6$ & $\Gamma_7$  & $\Gamma_8$ & $\Gamma_5$ & $\Gamma_4$ \\
  &   &   &   & $\Gamma_3$  &$\Gamma_1$ & $\Gamma_4$  & $\Gamma_2$ & $\Gamma_5$ \\
  &   &   &   &   & $\Gamma_4$ & $\Gamma_2$ & $\Gamma_3$  & $\Gamma_6$ \\
  &   &   &   &   &   & $\Gamma_3$  & $\Gamma_1$ & $\Gamma_7$  \\
  &   &   &   &   &   &   & $\Gamma_4$ & $\Gamma_8$
\end{tabular}
\label{SM:multS4}
\end{minipage}
\end{ruledtabular}
\end{table}

\setlength{\tabcolsep}{0.4em} 
\renewcommand{\arraystretch}{1.25}
\begin{table}[!htb]
\caption{Multiplication table for $D_{2d}$ symmetry \cite{Koster1963_S}.}
\begin{ruledtabular}
\begin{minipage}[t]{.6\linewidth}
\begin{tabular}{ccccccc|c}
$\Gamma_1$ & $\Gamma_2$ & $\Gamma_3$ & $\Gamma_4$  & $\Gamma_5$ & $\Gamma_6$ & $\Gamma_7$ &   \\ \hline
$\Gamma_1$ & $\Gamma_2$ & $\Gamma_3$  & $\Gamma_4$ & $\Gamma_5$ & $\Gamma_6$ & $\Gamma_7$  & $\Gamma_1$ \\
  & $\Gamma_1$ & $\Gamma_4$ & $\Gamma_3$  & $\Gamma_5$  & $\Gamma_6$ & $\Gamma_7$  & $\Gamma_2$ \\
  &   & $\Gamma_1$ & $\Gamma_2$ & $\Gamma_5$ & $\Gamma_7$ & $\Gamma_6$ & $\Gamma_3$  \\
  &   &   & $\Gamma_1$ & $\Gamma_5$ & $\Gamma_7$  & $\Gamma_6$ & $\Gamma_4$ \\
  &   &   &   & $\Gamma_1+\Gamma_2+\Gamma_3+\Gamma_4$  &$\Gamma_6+\Gamma_7$ & $\Gamma_6+\Gamma_7$  & $\Gamma_5$ \\
  &   &   &   &   & $\Gamma_1+\Gamma_2+\Gamma_5$ & $\Gamma_3+\Gamma_4+\Gamma_5$  & $\Gamma_6$ \\
  &   &   &   &   &   & $\Gamma_1+\Gamma_2+\Gamma_5$ & $\Gamma_7$  \\
\end{tabular}
\label{SM:multD2d}
\end{minipage}
\end{ruledtabular}
\end{table}

\setlength{\tabcolsep}{0.4em} 
\renewcommand{\arraystretch}{1.25}
\begin{table*}[!htb]
\caption{Observed selection rules for optical absorption between different hyperfine levels in ground \dfs{}(0) and \dfc{}(0) excited states. 
}
\begin{ruledtabular}
\begin{minipage}[t]{.45\linewidth}
\begin{tabular}{c|ccc}
       & $\ket{1,2}_e $  & $\ket{3}_e  $  & $\ket{4}_e $  \\ \hline
$\bra{1}_g$ &   $\alpha$, $\sigma$, $\pi$ & $\sigma$ & - \\
$\bra{2,3}_g $  & $\sigma$ & $\alpha$, $\sigma$, $\pi$ &  $\alpha$, $\sigma$, $\pi$   \\
$\bra{4}_e $  & $\alpha$, $\sigma$, $\pi$ &- & $\sigma$  
\end{tabular}
\end{minipage}%
\end{ruledtabular}
\label{SM:BRtab}
\end{table*}

\setlength{\tabcolsep}{0.4em} 
\renewcommand{\arraystretch}{1.25}
\begin{table*}[!htb]
\caption{ Selection rules for optical transitions between different hyperfine levels in ground and excited states for electric (black)  and magnetic  (\red{red}) dipoles for $S_4$ symmetry. Left: ground and excited states are assumed to have $\Gamma_{5,6}(\Gamma_{7,8})$ and $\Gamma_{7,8}(\Gamma_{5,6})$ irreps, respectively. Right: ground and excited states are assumed to have $\Gamma_{5,6} (\Gamma_{7,8})$ and $\Gamma_{5,6}(\Gamma_{7,8})$ irreps, respectively. }
\begin{ruledtabular}
\begin{minipage}[t]{.45\linewidth}
\begin{tabular}{c|ccc}
       & $\ket{1,2}_e (\Gamma_{3,4})$  & $\ket{3}_e  (\Gamma_{1})$  & $\ket{4}_e (\Gamma_{1})$  \\ \hline
$\bra{1}_g (\Gamma_{2})$ &   $\alpha$, $\sigma$ (\red{$\alpha$, $\pi$}) & $\pi$ (\red{$-$}) & $\pi$ (\red{$-$}) \\
$\bra{2,3}_g (\Gamma_{3,4})$  & $\pi$ (\red{$\sigma$}) & $\alpha$, $\sigma$ (\red{$\alpha$, $\pi$}) &  $\alpha$, $\sigma$ (\red{ $\alpha$, $\pi$})  \\
$\bra{4}_e (\Gamma_{2})$  & $\alpha$, $\sigma$ (\red{$\alpha$, $\pi$})  & $\pi$ (\red{$-$})  & $\pi$ (\red{$-$})
\end{tabular}
\end{minipage}%
\end{ruledtabular}
\begin{ruledtabular}
\begin{minipage}[t]{.45\linewidth}
\begin{tabular}{c|ccc}
       & $\ket{1,2}_e (\Gamma_{3,4})$  & $\ket{3}_e  (\Gamma_{2})$  & $\ket{4}_e (\Gamma_{2})$  \\ \hline
$\bra{1}_g (\Gamma_{2})$ &   $\alpha$, $\sigma$ (\red{$\alpha$, $\pi$}) & $-$ (\red{$\sigma$}) & $-$ (\red{$\sigma$}) \\
$\bra{2,3}_g (\Gamma_{3,4})$  & $\pi$ (\red{$\sigma$}) &  $\alpha$, $\sigma$ (\red{$\alpha$, $\pi$}) &  $\alpha$, $\sigma$ (\red{ $\alpha$, $\pi$})  \\
$\bra{4}_e (\Gamma_{2})$  & $\alpha$, $\sigma$ (\red{$\alpha$, $\pi$}) & $-$ (\red{$\sigma$}) & $-$ (\red{$\sigma$})
\end{tabular}
\end{minipage}%
\end{ruledtabular}
\label{SM:BRtabS4}
\end{table*}

\setlength{\tabcolsep}{0.4em} 
\renewcommand{\arraystretch}{1.25}
\begin{table*}[!htb]
\caption{ Selection rules for optical transitions between different hyperfine levels in ground and excited states for electric (black) and magnetic (\red{red}) dipoles   for $D_{2d}$ symmetry. Left: ground state is assumed to be $\Gamma_{6}(\Gamma_7)$ and excited state  $\Gamma_{7}(\Gamma_{6})$ irreps. Right: ground state is assumed to be $\Gamma_{7}(\Gamma_6)$ and excited state  $\Gamma_{7}(\Gamma_{6})$ irreps. }
\begin{ruledtabular}
\begin{minipage}[t]{.45\linewidth}
\begin{tabular}{c|ccc}
       & $\ket{1,2}_e (\Gamma_{5})$  & $\ket{3}_e  (\Gamma_{2})$  & $\ket{4}_e (\Gamma_{1})$  \\ \hline
$\bra{1}_g (\Gamma_{3})$ &   $\alpha$, $\sigma$ (\red{$\alpha$, $\pi$}) & $\pi$ (\red{$-$}) & $-$ (\red{$-$}) \\
$\bra{2,3}_g (\Gamma_{5})$  & $\pi$ (\red{$\sigma$}) & $\alpha$, $\sigma$ (\red{$\alpha$, $\pi$}) &  $\alpha$, $\sigma$ (\red{$\alpha$, $\pi$})  \\
$\bra{4}_e (\Gamma_{4})$  & $\alpha$, $\sigma$ (\red{$\alpha$, $\pi$})  & $-$ (\red{$-$})  & $\pi$ (\red{$-$})
\end{tabular}
\end{minipage}%
\end{ruledtabular}
\begin{ruledtabular}
\begin{minipage}[t]{.45\linewidth}
\begin{tabular}{c|ccc}
       & $\ket{1,2}_e (\Gamma_{5})$  & $\ket{3}_e  (\Gamma_{4})$  & $\ket{4}_e (\Gamma_{3})$  \\ \hline
$\bra{1}_g (\Gamma_{3})$ &   $\alpha$, $\sigma$ (\red{$\alpha$, $\pi$ })& $-$ (\red{$\sigma$}) & $-$ (\red{$-$}) \\
$\bra{2,3}_g (\Gamma_{5})$  & $\pi$ (\red{$\sigma$}) & $\alpha$, $\sigma$ (\red{$\alpha$, $\pi$}) &  $\alpha$, $\sigma$ (\red{$\alpha$, $\pi$})  \\
$\bra{4}_e (\Gamma_{4})$  & $\alpha$, $\sigma$ (\red{$\alpha$, $\pi$}) & $-$ (\red{ $-$})& $-$ (\red{$\sigma$}) 
\end{tabular}
\end{minipage}%
\end{ruledtabular}
\label{SM:BRtabD2d}
\end{table*}

\pagebreak
\clearpage
\begin{@fileswtrue}
\bibliographystyleS{apsrev4-1}
\bibliographyS{reflist_SM}
\end{@fileswtrue}

\end{document}